\documentclass[runningheads]{llncs}
\usepackage{subfigure}
\usepackage{fmtcount}
\usepackage{graphicx}
\usepackage{xspace}
\usepackage{url}
\usepackage{listings}
\usepackage{acronym}
\usepackage{latexsym}
\usepackage{fmtcount}
\usepackage{courier}
\usepackage{algpseudocode}
\usepackage[linesnumbered,ruled,vlined]{algorithm2e}

\begin{document}

\maketitle

\begin{abstract}
An adversary who aims to steal a black-box model repeatedly queries the model via a prediction API to learn a function that approximates its decision boundary. Adversarial approximation is non-trivial because of the enormous combinations of model architectures, parameters, and features to explore. In this context, the adversary resorts to a {\em best-effort strategy} that yields the closest approximation. This paper explores best-effort adversarial approximation of a black-box malware classifier in the {\em most challenging setting}, where the adversary's knowledge is limited to a prediction label for a given input. Beginning with a limited input set for the black-box classifier, we leverage {\em feature representation mapping} and {\em cross-domain transferability} to approximate a black-box malware classifier by locally training a substitute. Our approach approximates the target model with {\em different feature types} for the target and the substitute model while also using {\em non-overlapping data} for training the target, training the substitute, and the comparison of the two. We evaluate the effectiveness of our approach against two black-box classifiers trained on Windows Portable Executables (PEs).  Against a Convolutional Neural Network (CNN) trained on raw byte sequences of PEs, our approach achieves a 92\% accurate substitute (trained on pixel representations of PEs), and nearly 90\% prediction agreement between the target and the substitute model. Against a 97.8\% accurate gradient boosted decision tree trained on static PE features, our 91\% accurate substitute agrees with the black-box on 90\% of predictions, suggesting the strength of our purely black-box approximation.
\end{abstract}

\section{Introduction}
 Recent advances in machine learning (ML), specially in deep learning, have led to significant improvement on the accuracy of image classification, machine translation, speech processing, and malware/intrusion detection. Despite their impressive accuracy,  deep neural networks (DNNs) and other traditional machine learning models such as Logistic Regression, Support Vector Machines, and Decision Trees have been shown to be vulnerable to training-time poisoning~\cite{SVM-poisoning12,Wild-patterns18}, test-time evasion~\cite{Biggio-ECML13,Goodfellow14,Harness-adv-ex14,transferability16}, model extraction~\cite{model-stealing16,EfficientStealing19,KnockoffNets19}, and membership inference attacks~\cite{Membershp-inference17,Secret-sharer19}. 

With the advent of ML-as-a-Service (MLaaS), ML models are increasingly served via prediction APIs to allow remote submission of input samples to produce predictions or provide pre-trained models as foundations to build up on.
While MLaaS enables new frontiers of ML use-cases such as serving models from the cloud with pay-per-query price model, it also exposes models behind prediction APIs to model extraction/approximation attacks via iterative input-output interactions \cite{model-stealing16,Membershp-inference17,PRADA19,EfficientStealing19,KnockoffNets19}. An adversary aiming to game a prediction API of a model to avoid a pay-per-prediction bill  or a competitor targeting the trade secret of a model have financial motivations to steal ML models. For instance, adversaries are motivated to steal (approximate) a remote black-box model trained on privacy-sensitive data (e.g., medical records), intellectual property (e.g., stock market trends), or inherently security-sensitive data (e.g., malware/intrusion traces). In general, so long as the cost of approximating an ML model is lower than the potential financial gain from obtaining a close-enough copy of it, MLaaS will continue to be a target of financially motivated adversaries.

More precisely, given a black-box model $f_{b}$ (e.g., a malware detector) served via a prediction API, the adversary's goal is to perform {\em best-effort} approximation of $f_{b}$'s decision boundary by locally training a substitute model $f_{s}$. Best-effort in this sense refers to relying on limited seed-set (e.g., 5\%-10\% of the training set for $f_{b}$) to probe $f_{b}$ and leveraging publicly accessible resources (e.g., features, pre-trained models) for effective and efficient approximation of $f_{b}$.

Previous work explored black-box model approximation by leveraging the fidelity (e.g., probability scores) of predictions~\cite{model-stealing16,PRADA19}, feature and/or model architecture similarity between $f_{b}$ and $f_{s}$ ~\cite{MalGAN17,model-stealing16,transferability16,EfficientStealing19}, and cross-model transferability ~\cite{transferability16,PracticalBBAtk}. The approximation formulation spans equation solving \cite{model-stealing16}, optimization methods \cite{HighExtr20}, generative adversarial networks~\cite{MalGAN17}, and reinforcement learning~\cite{KnockoffNets19}.

This paper explores adversarial approximation of a black-box malware detector $f_{b}$ in the {\em most challenging setting for the adversary}. In particular, we explore a threat model where the adversary aims for a close-enough approximation of $f_{b}$ in the face of $(i)$ access to limited inputs to $f_{b}$, $(ii)$ for a given input sample no additional observations beyond prediction label, $(iii)$ different feature representations for $f_{b}$ and $f_{s}$, and $(iv)$ disjoint training sets for $f_{b}$, $f_{s}$, and the similarity evaluation set. To that end, beginning with limited seed-set for the black-box classifier, we leverage {\em representation mapping} and {\em cross-domain transferability} to approximate a black-box malware classifier by locally training a substitute.  

Our work complements prior work~\cite{MalGAN17,model-stealing16,transferability16,PRADA19,EfficientStealing19,PracticalBBAtk} in three ways. First,  we do not assume any adversarial knowledge other than prediction label for a given PE. This is a strong adversarial setting, which effectively leaves the adversary no leverage except a best-effort strategy that relies on publicly available vantage points (e.g., input samples, pre-trained models). Second, we approximate $f_{b}$ with {\em different feature types} for $f_{b}$ (e.g., byte sequences) and $f_{s}$ (e.g., pixel intensities). By mapping the representation of PEs from byte-space to pixel-space, our approach eliminates  the need for manual feature engineering. The motivation behind using dissimilar feature representations for $f_{b}$ and $f_{s}$ is to leverage publicly accessible pre-trained models (e.g., Inception V3~\cite{inceptionv3}) by means of transfer learning, while only training on the last layer. While prior work \cite{transferability16} demonstrated cross-model transferability for image classifiers, we show a different dimension of transferability in a cross-domain setting by re-purposing a pre-trained image classifier to approximate a black-box malware detection model trained on raw-byte sequences of PEs.  Third, we deliberately use {\em non-overlapping data} for training $f_{b}$, training $f_{s}$, and comparison of similarity between the two. We note that some prior work~\cite{model-stealing16,transferability16,PracticalBBAtk,KnockoffNets19,PRADA19} use disjoint data when $f_{b}$ is hosted by MLaaS providers. It is, however,  hard to verify the disjointedness of $f_{b}$'s training data against $f_{s}$'s or the comparison set, because in such a setting $f_{b}$'s training data is typically confidential.  

Against a black-box CNN \cite{malconv18} trained on byte sequence features of Windows PEs, our approximation approach obtained up to 92\% accurate CNN on pixel features, and trained based on the Inception V3 \cite{inceptionv3} pre-trained model (details in \ref{subsec: approximation-results}). On a comparison dataset disjoint with the black-box's and the substitute's training sets, our approach achieved nearly 90\% similarity between the black-box CNN and the substitute one. 
In a nutshell, the results suggest that, even if the target model is a black-box, an adversary may take advantage of a limited training data and the underlying knowledge of pre-trained models (Inception V3 in this case) to successfully approximate the decision boundary of a black-box model.
An intriguing observation of our results is that, although the training samples used for approximation are disjoint with training samples used to train the black-box CNN, the adversary can still achieve an acceptable approximation of the black-box CNN with minimal efforts. Another intriguing observation is, despite the dissimilarity of the representation of the black-box (i.e., byte sequences) and that of the substitute model (i.e., pixels), our approximation approach still managed to achieve nearly 90\% similarity between the target black-box and the substitute model. 

We further evaluate our approach on a research benchmark dataset, EMBER~\cite{ember18}, which is based on static PE features (details in \ref{subsec: similarity-results}). Our approach approximated the LightGBM \cite{LGBM} black-box model supplied with EMBER via a 91\% accurate substitute model, which agrees with the black-box on 90\% of test PEs. With EMBER, we also explore how our approximation strategy performs on different models by training multiple substitute models, confirming the validity of our approach across model architectures such as Decision Trees, Random Forests, and K-Nearest Neighbours. In summary, this paper makes the following contributions:

\begin{itemize}
 \item By mapping byte sequence features to pixel representations, we eliminate the need for heuristics-based feature engineering, which significantly reduces adversarial effort towards feature guessing.
\item Leveraging a pre-trained model as a foundation, we extend the scope of transferability from a cross-model setting by demonstrating the utility of cross-domain transferability for black-box approximation.
\item We demonstrate the feasibility of close-enough adversarial approximation using different feature representations for the black-box and the substitute model, across multiple model architectures, and complementary datasets.
\end{itemize}

\section{Background and Threat Model}

\subsection{Model Approximation Attacks}
In this work, we focus on supervised learning for malware classification models, where input samples are Windows PEs and the output is the class label, i.e., {\tt benign} or {\tt malware}.

Let $X$ be a $d$-dimensional feature space and $Y$ be the $c$-dimensional output space, with underlying probability distribution $Pr(X,Y)$, where $X$ and $Y$ are random variables for the feature vectors and the classes of data, respectively. The objective of training an ML model is to learn a parameter vector $\theta$, which represents a mapping $f_{\theta} : X \rightarrow Y$. $f_{\theta}$ outputs a $c$-dimensional vector with each dimension representing the probability of input belonging to the corresponding
class. $l(f_{\theta}(x),y)$ is a loss of $f_{\theta}$ on $(x,y)$, and it measures how ``mistaken" the prediction $f_{\theta}(x)$ is with respect to the true label $y$. Given a set of training samples $D_{train} \subset (X,Y)$, the objective of
an ML model, $f_{\theta}$, is to minimize the expected loss over all
$(x,y): L_{D_{train}}(f_{\theta}) = \sum_{ (x,y) \in D_{train}}^{}l(f_{\theta}(x),y)$. In ML models such as DNNs, the loss minimization problem is typically solved using Stochastic Gradient Descent (SGD) by iteratively updating the weights $\theta$ as:  $\theta \leftarrow \theta - \epsilon\cdot \Delta_{\theta}(\sum_{(x,y) \in D_{train}} l(f_{\theta}(x),y))$, where $\Delta_{\theta}$ is the gradient of the loss with respect to the weights $\theta$; $D_{train} \subset (X,Y)$ is a randomly selected set ({\em mini-batch}) of training examples drawn from $X$; and $\epsilon$ is the {\em learning rate} which controls the magnitude of change on $\theta$.

\begin{figure*}[htb!]
    \centering
    \includegraphics[scale=1.1]{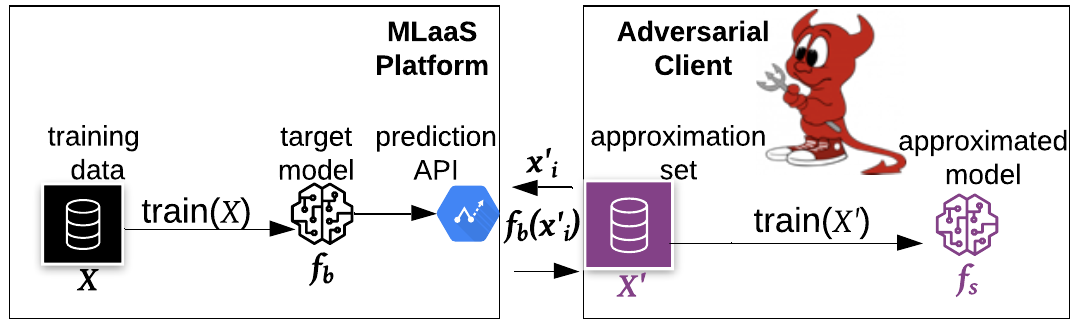}
    \caption{Typical pipeline for model approximation attacks. MLaaS provider trains a model $f_{b}$ on confidential/proprietary data $X$. It then serves $f_{b}$ via a prediction API to allow clients to issue queries and obtain prediction results. An adversarial client uses $f_{b}$ as an oracle to label $|X'|$ samples to train $f_{s}$ such that $f_{s} \approx f_{b}$. }
    \label{fig:approx-attack}
\end{figure*}

Figure~\ref{fig:approx-attack} depicts a typical setup for model approximation attacks against a MLaaS platform that serves a black-box model $f_{b}$ via a prediction API. The goal of the adversarial client is to learn a close-enough approximation of $f_{b}$ using as few queries ($x'_{i}$'s) as possible.

 Typically, model approximation is used to steal a deployed model (e.g., intellectual property, security-sensitive business logic) \cite{model-stealing16,HighExtr20,EfficientStealing19}, trick/bypass pay-per-prediction walls, perform reconnaissance for later attacks that go after the integrity of the victim model through adversarial examples \cite{transferability16}, or mount membership inference attacks for models known to be trained on privacy-sensitive data \cite{Membershp-inference17,Secret-sharer19}.

The success of approximation depends on the adversarial client's knowledge (about details of $f_{b}$ and its training set) and capabilities (e.g., access to inputs, maximum number of queries). In a {\em white-box} (full knowledge) setting, the adversary has complete knowledge of the target model architecture and its training data, making it the easiest for the adversary (but the most exposing for the target model). In a {\em grey-box} (partial knowledge) setting, the adversary has some knowledge about $f_{b}$'s model architecture (e.g., DNN vs. SVM) and training data (e.g., a subset of the training set), which makes it moderately challenging. In a {\em black-box} (no/limited knowledge) setting, the adversary knows nothing but prediction labels (and possibly confidence probabilities) for a given input. This is the most challenging setting for the adversary. In this paper, we consider the black-box setting. Next, we describe our threat model.

\subsection{Threat Model and Problem Statement}\label{sec:problem}

We consider the strongest threat model compared with previous work on adversarial approximation \cite{model-stealing16,EfficientStealing19,transferability16}. The adversary interacts with a deployed model $f_{b}$, for instance served from MLaaS malware detector. Without loss of generality, we assume that the ML model $f_{b}$ is a malware detector.

\textbf{Adversary's Goals:} The adversary's goal is to approximate the decision boundary of $f_{b}$ by training its substitute $f_{s}$, beginning with a limited seed-set of PEs.

\textbf{Adversary's Knowledge:}  The adversary only knows that $f_{b}$ accepts Windows PEs as input and returns labels (``benign'' or ``malicious'') as output. The adversary doesn't know $f_{b}$'s architecture, parameters, hyper-parameters, or features.  Besides, the adversary has no access to the training data or the test data used to train and evaluate $f_{b}$.

\textbf{Adversary's Capabilities:} The adversary probes $f_{b}$ with PEs to obtain prediction labels. We assume there is an upper bound on the number of queries the adversary can issue to $f_{b}$, but the adversary can workaround the query limit by probing $f_{b}$ over an extended time-window or from different IP addresses.  The adversary has access to a limited seed-set of PEs, but is unable to tell whether or not the PEs in the seed-set overlap with $f_{b}$'s training set. The adversary is able to continuously collect more PEs to use $f_{b}$ as an oracle and progressively train $f_{s}$ with the goal of obtaining a close-enough approximation of $f_{b}$. 

\textbf{Problem Statement:}
 Given a deployed malware detection model, $f_{b}$, under the threat model stated earlier, the adversary's goal is to find $f_{b}$'s closest approximation, $f_{s}$, with best effort. By ``best effort'' we mean relying on limited seed-set to probe $f_{b}$, and leveraging publicly accessible resources (e.g., feature representations, pre-trained models) towards effective and efficient approximation of $f_{b}$. 
 
 More formally, given a black-box model $f_{b}$ trained on dataset $X$, for a seed-set $|X^\prime| < |X|$  and $X^\prime \cap X = \emptyset$, the adversary's goal is to train $f_{s}$ using $X^\prime$ such that when compared on  dataset $X^{\prime\prime}$ disjoint with $X$ and $X^\prime$, $f_{b} \approx  f_{s}$. The $\approx$ quantifies the percentage of agreement (i.e., number of matching predictions between $f_{b}$ and $f_{s}$ $X^{\prime\prime}$). The closer the agreement is to a 100\% the more accurate the approximation and vice-versa. The real-life implication of successful approximation of $f_{b}$ via $f_{s}$ is that once the adversary obtains a close-enough (e.g., $>90\%$) substitute of $f_{b}$, the intellectual property of $f_{b}$'s owner (e.g., an IDS vendor) is jeopardized. Even worse, the adversary might as well emerge as $f_{b}$'s competitor by tuning $f_{s}$'s accuracy with additional training data.
\section{Approach}

\begin{figure*}[htb!]
    \centering
    \includegraphics[scale=.55]{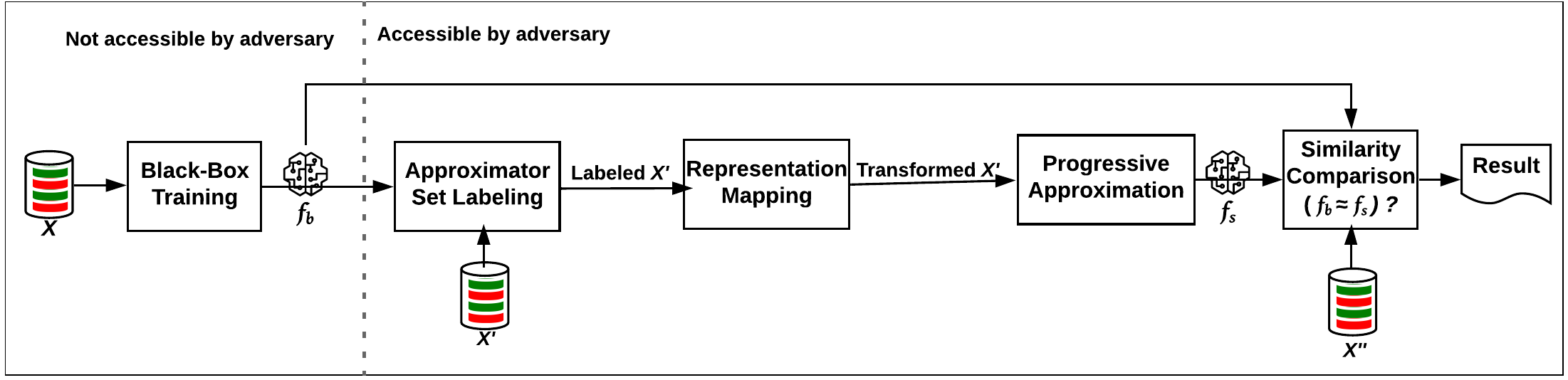}
    \caption{Approach overview. $f_{b}$ is accessible by the adversary via a prediction API only.}
    \label{fig:approach}
\end{figure*}

\textbf{Overview:} Figure \ref{fig:approach} depicts an overview of our approach. Given a black-box malware detector $f_{b}$, the adversary collects benign and malware PEs from public sources to obtain the substitute training set ($X^\prime$). The adversary then uses $f_{b}$ as an oracle to label samples in $X^\prime$ (Section \ref{subsec: label}). Next, samples in $X^\prime$ are mapped from a source representation: raw-bytes to a target representation: pixel intensities (Section. \ref{subsec: transform}). The progressive approximation step uses the mapped  $X^\prime$ to iteratively approximate $f_{b}$ (Section \ref{subsec: approximate}). It combines {\em representation mapping} and {\em cross-domain transferability} to obtain a close-enough approximation of $f_{b}$ with a limited seed-set to $f_{b}$ (i.e.,  $|X^\prime| < |X|$), different feature representations for $f_{b}$ and $f_{s}$, and disjoint training sets for $f_{b}$, $f_{s}$, and the comparison set used to evaluate similarity between $f_{b}$ and $f_{s}$ (i.e., $X \cap X^\prime \cap X^{\prime\prime} = \emptyset$). The approximation begins by training an initial substitute model $f_{s}$ on a fraction (e.g., 20\%-25\%) of $X^\prime$, and $f_{s}$ is refined until it achieves acceptable accuracy.
Using a dataset $X^{\prime\prime}$, the last stage of the approach compares similarity between $f_{b}$ and $f_{s}$ (Section \ref{subsec: compare}). A higher similarity score for $f_{b}$ and $f_{s}$ is an indication of the effectiveness of the approach at approximating the decision boundary of $f_{b}$. 

For the sake of presenting an end-to-end approximation framework, Figure \ref{fig:approach} includes $f_{b}$'s training at the beginning. In practice, the adversary doesn't have access to the black-box model except through its prediction API. It is also noteworthy that the three datasets ($X$, $X^\prime$, and $X^{\prime\prime}$) shown in Figure \ref{fig:approach} are disjoint. Again, in reality, with no access to $f_{b}$'s training set, the adversary has no trivial way to determine if the substitute training set ($X^\prime$) or the model similarity comparison set ($X^{\prime\prime}$) have intersection among themselves or with the black-box training set ($X$). The only motivation of training our own black-box model $f_{b}$ is to ensure $X$, $X'$, and $X''$  are disjoint, and doing  so lets us to deterministically examine the most challenging adversarial setting described earlier.

\subsection{Approximation Set Labeling} \label{subsec: label}
Given a seed-set of Windows PEs collected by the adversary, a first-cut labeling strategy would be to take the ground truth labels that the samples come with (e.g., using VirusTotal \cite{virustotal} as an oracle). Had our goal been to train a malware detector, such an approach would suffice. Our goal, however, is to approximate the decision boundary of a black-box malware detector, and we expect our labeling method to serve this purpose. 

Given a set of approximation samples $ X' = x'_{1}, ..., x'_{m}$ and a learned hypothesis function $f_{b}$,  we obtain $f_{b}(x'_{i}) = y'_{i}$. The $y'_{i}$'s may or may not match the ground truth labels. If $f_{b}$ misclassifies some $x'_{i}$'s,  the misclassified  $x'_{i}$'s will not match the ground truth counter-parts. What should be done with the misclassified samples in the substitute training set? The alternatives we have are (a) drop the misclassified  $x'_{i}$'s and explore approximation with the correctly labeled $x'_{i}$'s,  (b) reinstate labels to ground truth labels and proceed with approximation, or (c) take the labels assigned by $f_{b}$ for what they are. Alternative (a) is no different from training the substitute without querying $f_{b}$. Alternative (b) entails ``correcting'' the ``imperfections'' of $f_{b}$ (note ``correcting'' could as well mean lower accuracy because we are essentially changing the underlying distribution of $f_{b}$'s training set). Alternative (c) is the most realistic for our threat model, because it takes $f_{b}$ for what it is and uses its predictions ($y'_{i}$'s) to populate the labeled approximation set, which is highly likely to result in a realistic approximation of $f_{b}$'s decision boundary.

\subsection{Representation Mapping}\label{subsec: transform}
Under our threat model, the adversary doesn't know what features are used to train $f_{b}$.  The adversary may then pursue different possibilities of features used to train malware detectors based on Windows PEs. However, the space of possible features is exponentially large. For example, if we only consider static analysis based features of a given PE, we end up with numerous possible features such as meta-data, DLL imports/exports, byte sequences, and so on. Similarly, considering the dynamic analysis-based features of PEs results in several candidates such as API/system call traces, instruction sequences, and call graphs. Therefore, given enough resources, while such a strategy of feature guessing may result in a close-enough approximation, it may not be the preferred avenue by the adversary whose goal is to effectively and efficiently  approximate $f_{b}$.

In-line with the adversary's goals, we map the raw bytes representation of each PE to an image representation (pixel values), analogous to taking a photograph of the PE's raw byte sequences. The main rationale is, instead of searching for the best combination of a set of features to train the substitute, it is plausible to capture the whole PE's bytes via image representations such that the learning algorithm is able to ``discover'' distinguishing sub-structures from the image representation of PEs. We note that previous work (\cite{malware-images11,malware-entropy15,copycat19}) has also explored bytes-to-pixels conversion for malware detection, although not in the exact context of model approximation via cross-domain transferability that we explore in this work. Another equally important rationale for bytes-to-pixels mapping is the accessibility (to the adversary) of acceptably accurate pre-trained image classification models such as Inception V3 \cite{inceptionv3} which, by way of transfer learning \cite{transferability16}, feed knowledge to the substitute model.
\begin{figure*}[htb!]
    \centering
        \includegraphics[width=\textwidth]{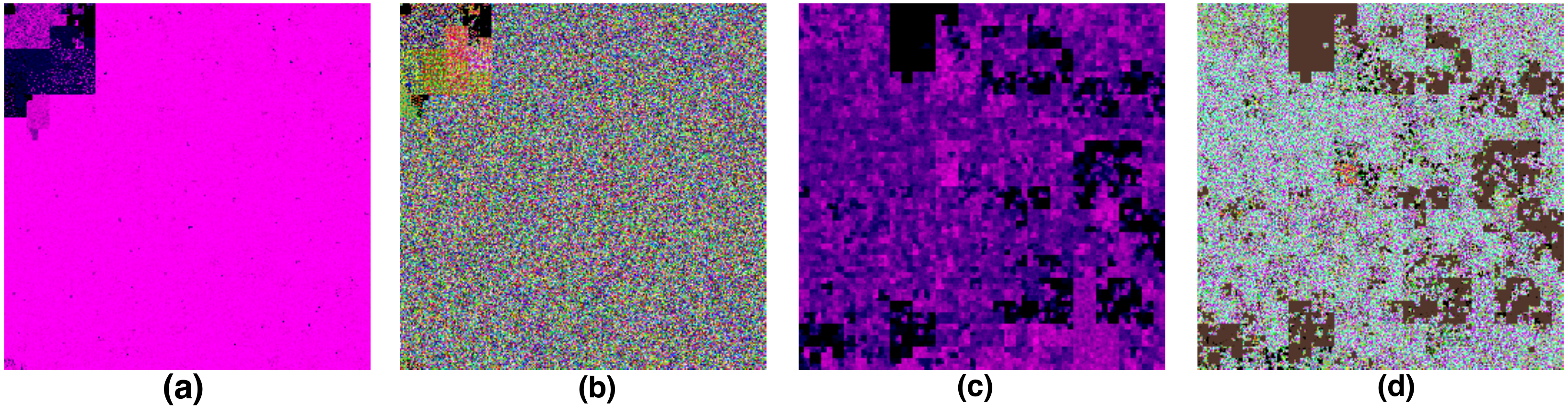}
        \caption{(a) and (b) show EN and CH rendering, respectively, of a benign PE (small02Micro\_Card\_Reader\_Driver\_3.11.exe). (c) and (d) show EN and CH rendering, respectively, of a  malware PE (Trojan.GenericKDZ.58985).}
    \label{fig: en-ch-demo}
\end{figure*}

To realize the bytes-to-pixels mapping, the key intuition is to transform PEs into a colored canvas, such that the colors (pixel intensities) represent the bytes in the PEs, and the color intensities are used as features to train the substitute. To this end, we leverage two types of image representations, the Entropy (EN) representation \cite{shanon} and the Color Hilbert (CH) representation \cite{binviz}. CH scans the bytes of a PE and assigns color based on the value of each byte. The assigned pixel intensities are then mapped on a canvas of a chosen dimension. EN uses Shannon Entropy ~\cite{shanon} to compute the randomness of bytes in a specific location of the executable as $E = -\sum_{i=1}^{n}\rho_{i} \log_{2} \rho_{i}$,  where $\rho_{i}$ refers to the probability of appearances of a byte value $i$ and $n$ is the number of possible values ($n = 256$ for possible byte values). Depending on the computed value of $E$, the corresponding pixel is assigned color intensity in the range black (minimum/zero entropy) to bright pink (maximum entropy). An example that illustrates EN and CH representation for a benign PE is shown in Figure \ref{fig: en-ch-demo} (a) and (b), respectively. Similarly, Figure \ref{fig: en-ch-demo} (c) and (d) show EN and CH representations of a malware PE, respectively. Notice the clear visual difference between benign and malware PEs, which seems to support our intuition of mapping bytes to pixels. Looking at the CH representation, the combination of these colors in the images serves best to give the model discriminating features to put apart benign and malware PEs. With regards to the EN representation, we can see that the focus is more on areas instead of specific pixels, and it is not as high-fidelity as the CH representation. In section \ref{chap:results}, we evaluate the utility of EN and CH representations via the substitute's accuracy. Next, we briefly describe how the canvas is filled with colors.

To paint the canvas of the image, we leverage a well-known mapping technique called the Hilbert curve~\cite{hilbert} (implemented in BinVis~\cite{binviz}), which makes sure that if two bytes are close to each other in the PE, they should be close to each other in the image representation of the PE as well. This property of the Hilbert curve is essential to preserve the semantic structure of PEs when mapping bytes to pixels, and it provides the substitute model an accurate representation of the PE so that, during training, it explores the classification utility of all possible features. 
A natural question would be, how does the Hilbert curve function? Intuitively, the idea of Hilbert curve is to find a line to fill a canvas that will keep the points which are close to each other on that line at the same distance when it fills the needed space. This, in our case, keeps the features of the PEs intact since separating them would lead to breaking the semantics of the sequence of bytes that represents a part of our feature set in the images we would like to generate at the end of the mapping. 

Note that although representation mapping is core to our approach, sometimes, feature guessing may fit the best-effort strategy when the adversary has  access to publicly released datasets such as EMBER~\cite{ember18}, which reveal details about features. To account for this possibility, in Sections~\ref{subsec: approximation-results} and \ref{subsec: similarity-results}, we evaluate our approach with minimal relaxation on our threat model, i.e., the adversary knows features because of public disclosure as in~\cite{ember18}. As we will show, the purely black-box approximation is as strong as the minimally relaxed approximation.

\subsection{Progressive Approximation}\label{subsec: approximate}
On the one hand, the adversary has access to limited  input samples to the black-box model. On the other hand, the adversary has access to pre-trained and publicly  accessible image classification models (e.g., Inception V3 \cite{inceptionv3}). To obtain an acceptably accurate approximation of the black-box model, the adversary takes advantage of the advances in image classification models to quickly and accurately train a candidate substitute on the mapped features of the PEs via {\em cross-domain transferability}. The motivation behind leveraging pre-trained models is threefold. First, the fact that our substitute model relies on image classification for which state-of-the-art benchmark datasets are readily accessible. Second, the prospect of using a widely available and acceptably accurate pre-trained model for anyone (including the adversary) is in favor of the constraints the adversary has on collecting enough training samples for the approximation and come up with an effective model architecture tuned to adversarial goals. Third, when using pre-trained models, we are only retraining the last layer, which not only saves us (the adversary) time, but also gives us confidence on the accuracy of the final model because of transferability. Moreover, taking advantage of image representations cuts the effort on feature engineering down to zero because the candidate substitute (e.g., a CNN) would automatically learn the features from its mapped training data, i.e., image representations of PEs.

% \begin{algorithm}[h!]
% \SetAlgoLined
% $\tau_{acc}$: accuracy threshold\;
% $\tau_{sim}$: similarity threshold\;
% $num\_batches$: number of approximation set batches\;
%  \For {i =1 \to num\_batches}{
%   $acc_{i}, f_{s} \gets TrainSubstitute(f_{s}, batch_{i})$\;
%   $sim_{i} \gets GetSimilarity(f_{b}, f_{s})$\;
%   \If {acc_{i} > \tau_{acc}  \&\&  sim_{i} > \tau_{sim}}{
%   $StopApproximation()$\;
%   }
%  }
% \caption{Progressive approximation of $f_b$.}
% \label{alg: approx}
% \end{algorithm}

\begin{figure*}[htb!]
    \centering
    \includegraphics[scale=.4]{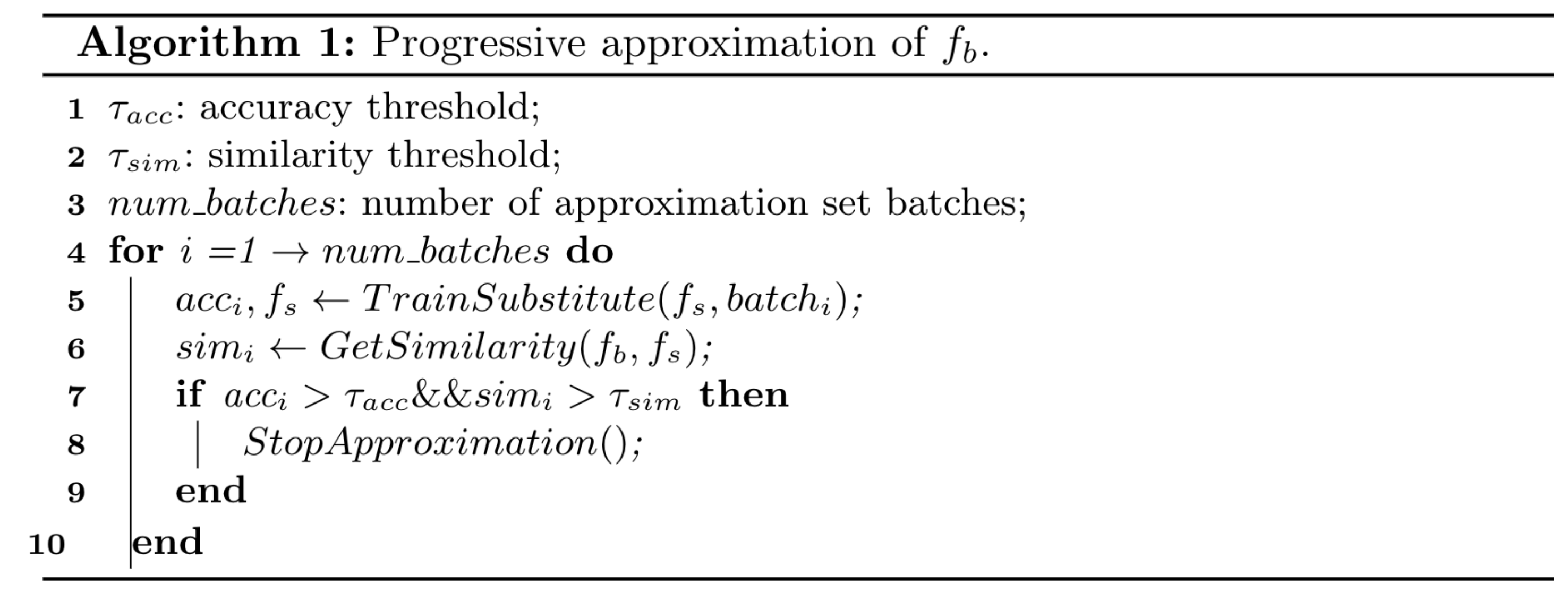}
    \caption{Progressive approximation of $f_b$.}
    \label{fig:alg1}
\end{figure*}

In order to emulate an adversary who begins with a limited seed-set to bootstrap approximation, we assume that the adversary has $batch_{1}$ as the first batch (e.g., 20\%) of substitute training samples $X'$. The adversary first trains $f_{s}$ on $batch_{1}$ and evaluates its accuracy against a pre-determined threshold, which could be set close-enough to $f_{b}$'s accuracy for our case. In real-life, one can safely assume that the adversary would estimate the accuracy threshold from public knowledge (e.g, state of the art model accuracy for malware classifiers). Similarly, the adversary can set a similarity threshold based on prediction matches between $f_{b}$ and $f_{s}$ over dataset $X^{\prime\prime}$. The adversary would then actively collect training data, and progressively re-train and re-evaluate $f_{s}$ each time they obtain the next batch of approximation examples, until an acceptable accuracy and similarity score is obtained, or the training data is exhausted. Figure \ref{fig:alg1} shows the details of the progressive approximation.

We note that previous work, specially on image classification models, relies on {\em data augmentation}  \cite{transferability16,PracticalBBAtk} techniques to address data scarcity for training a substitute.  Augmentation  aims to synthesize minimally manipulated, yet semantically intact variants of samples (e.g., images). Although augmentation techniques such as flipping and rotation would yield more synthetic samples for training a substitute, in contrast to image classifiers, minor flipping or rotation may result in broken semantics in PEs. To demonstrate the utility of augmentation, we explored how to extend our approximation dataset by three-fold via minor flipping, rotation, and flipping followed by rotation (details in Appendix A).

\subsection{Similarity Comparison} \label{subsec: compare}
\begin{figure*}[htb!]
    \centering
    \includegraphics[scale=.4]{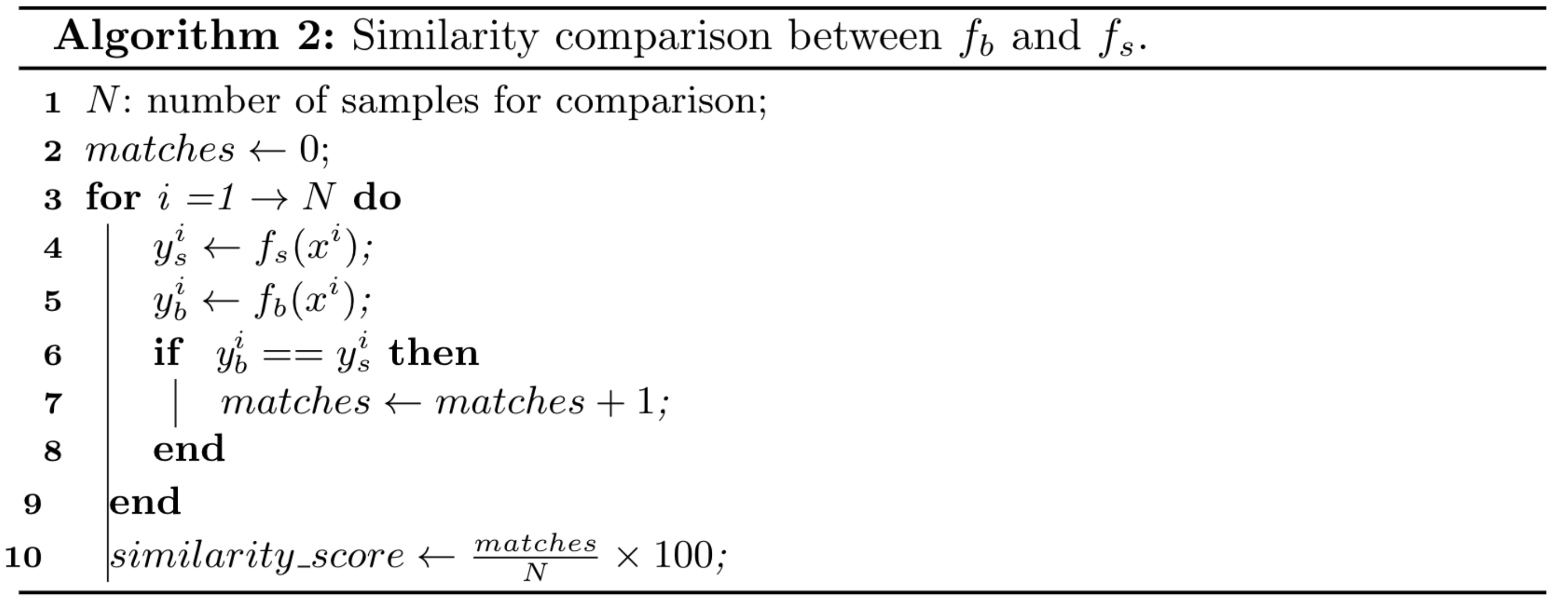}
    \caption{Similarity comparison between  $f_b$ and $f_s$.}
    \label{fig:alg2}
\end{figure*}

Once the substitute model $f_{s}$ is trained with an acceptable accuracy, its effectiveness is assessed when compared with the black-box model on a separate dataset, disjoint with both the training set of the black-box and the substitute.  Figure \ref{fig:alg2} shows the procedure for similarity comparison of $f_{b}$ and $f_{s}$.

% \begin{algorithm}[h!]
% %\scriptsize
% \SetAlgoLined
% %\KwResult{EvasionRate}
% %\textbf{Data:}
% $N$: number of samples for comparison\;
% $matches \gets 0$\;
%   \For {i =1 \to N}{
%   $y^{i}_{s} \gets f_s(x^{i})$\;
%     $y^{i}_b \gets f_b(x^{i})$\;
%   \If { y^{i}_{b} == y^{i}_{s} }{
%   $matches \gets matches +1$\;
%   }
% }
% $similarity\_score \gets \frac{matches}{N} \times 100 $\;
% \caption{Similarity comparison between  $f_b$ and $f_s$.}
% \label{alg: sim-score}
% \end{algorithm}

The similarity score is the percentage of matching predictions between $f_{b}$ and $f_{s}$. The higher the similarity score, the closer $f_{s}$ is to $f_{b}$, which means $f_{s}$ effectively mirrors the decision boundary of $f_{b}$. The adversary then probes $f_{s}$ for further attacks in a white-box setting. Attacks that succeed on $f_{s}$, would, by transitivity, succeed on $f_{b}$. This is the essence of having an accurate approximation which would be used as a substitute for the black-box. By crafting adversarial examples using methods such as FGSM \cite{Goodfellow14} and CW \cite{Carlini-Wagner17}, the adversary can transitively target $f_{b}$ using $f_{s}$ as a surrogate white-box model. Comparison of our candidate substitutes with the black-box model is discussed in Section \ref{subsec: similarity-results}.

% !TEX root =  main.tex
\section{Evaluation}   \label{chap:results}
We now describe our datasets, experimental setup, and results on progressive approximation and similarity comparison.

\subsection{Datasets}\label{subsec: dataset}
Table \ref{tab: dataset} summarizes our datasets:  Custom-MalConv and EMBER\cite{ember18}.

\textbf{Custom-MalConv:}   We collected benign PEs from a Windows freeware site \cite{cnet} and malware PEs from VirusShare \cite{virusshare}. These two sources are used by prior work \cite{malconv18,ember18,SLEIPINIR18,MalGAN17,APIBlackBoxEvade18} on adversarial examples and malware detection. Overall, we collect 67.81K PEs with 44\% benign and 56\% malware. We use 60\% for training the black-box CNN, 23\% as substitute training set, and 17\% as similarity comparison set. Since we control the dataset,  we use this dataset to evaluate the effectiveness of our approach on representation mapping and cross-domain transferability.

\textbf{EMBER:} EMBER~\cite{ember18} is a public dataset of malware and benign PEs released together with a Light gradient boosted decision tree model (LGBM) \cite{LGBM} with 97.3\% detection accuracy. The dataset consists of 2351 features extracted from 1M PEs via static binary analysis. The training set contains 800K labeled samples with 50\% split between benign and malicious PEs, while the test set consists of 200K samples, again with the same label split. The authors of EMBER used VirusTotal~\cite{virustotal} as a labeling oracle to label samples in EMBER. We use this dataset to further evaluate the generalizability of our approach. We note, however, that EMBER releases only static analysis based features extracted from PEs, not the PEs themselves, which limits our scope of doing representation mapping. Despite the absence of original PEs, we still evaluate our approach on a different feature space and over multiple candidate substitute models. Of the 200K test set samples, we use 150K as substitution training set, and the remaining 50K (25\%) as similarity comparison set. 
\begin{table}[t!]
 %\scriptsize
    \centering
\begin{tabular}{|l|r|r|r|}
\hline \textbf{Dataset} & \textbf{Benign}  & \textbf{Malware}& \textbf{Total}\\ \hline 
        \textbf{Custom-MalConv} &&&\\ \hline

    Black-Box training set & 20, 000 & 20, 000 & 40, 000\\  
    Substitute training set  & 8, 000 & 8, 000 & 16, 000\\ 
    Similarity comparison set & 2, 045 & 9, 765 & 11, 810\\
    \textbf{Total} & & & \textbf{67, 810}\\ \hline 
    \textbf{EMBER}\cite{ember18}& &&\\ \hline
    Black-Box training set & 400, 000 & 400, 000 & 800, 000\\ 
    Substitute training set  & 75, 000 & 75, 000 & 150, 000\\ 
    Similarity comparison set & 25, 000 & 25, 000 & 50, 000\\
    \textbf{Total} & & & \textbf{1, 000, 000}\\ \hline
\end{tabular}
\caption{Datasets for Custom-MalConv and EMBER.}
\label{tab: dataset}
\end{table}
\vspace{-.25in}

\subsection{Experimental Setup}\label{subsec:setup}

\textbf{Black-box Models:} The Custom-MalConv black-box is trained on 20K benign and 20K malware PEs. It is based on MalConv \cite{malconv18}, a widely used malware detection CNN in adversarial ML literature on malware \cite{MalConvEvade18,ExploreAdvEx18}. We use the same architecture as the original MalConv~\cite{malconv18} with slight modifications to fit our hardware (NVIDIA 1080 with 8GB of RAM). MalConv is based on raw bytes of a PE and reads the first 1MB to extract byte features. To fit our hardware limitations, we fed only $\frac{1}{3}$MB of each PE to our custom CNN, and got an accuracy of 93\% which is acceptable since MalConv \cite{malconv18} used 100K PEs to achieve 98\% accuracy. The LightGBM black-box is trained on 400K benign and 400K malware PEs and is shipped with the EMBER~\cite{ember18} dataset with 97.3\% accuracy. 

\textbf{Substitute Models:}
For Custom-MalConv black-box model, our $f_{s}$ is based on Inception V3 (IV3) \cite{inceptionv3} and is trained on both CH and EN representations of the substitute training set. In addition, for sanity check, we trained a substitute Custom-MalConv model with exact same features and model parameters as our black-box. For EMBER, we explore 4 candidates for substitute model, namely: decision trees (DT), random forests (RF), $k$-Nearest neighbors ($k$NN), and gradient-boosted decision tree model (LGBM).

\textbf{Progressive Approximation:}
For Custom-MalConv, we use 16K substitute training set and train $f_{s}$ progressively on 4K, 8K, 12K, and 16K PEs. For LGBM, we use 150K of the 200K test set with a progression of 30K, 60K, 90K, 120K, and 150K PEs. The metrics we use are {\em validation accuracy} and {\em similarity score}. Validation accuracy is the prediction accuracy of the approximated substitute. Similarity score (as shown in Figure \ref{fig:alg2}) is the percentage of matching predictions between $f_{b}$ and $f_{s}$.

\subsection{Progressive Approximation Results}\label{subsec: approximation-results}

\begin{figure*}[t!]
\begin{minipage}[b]{0.5\linewidth}
 \centering
 \includegraphics[scale=.38]{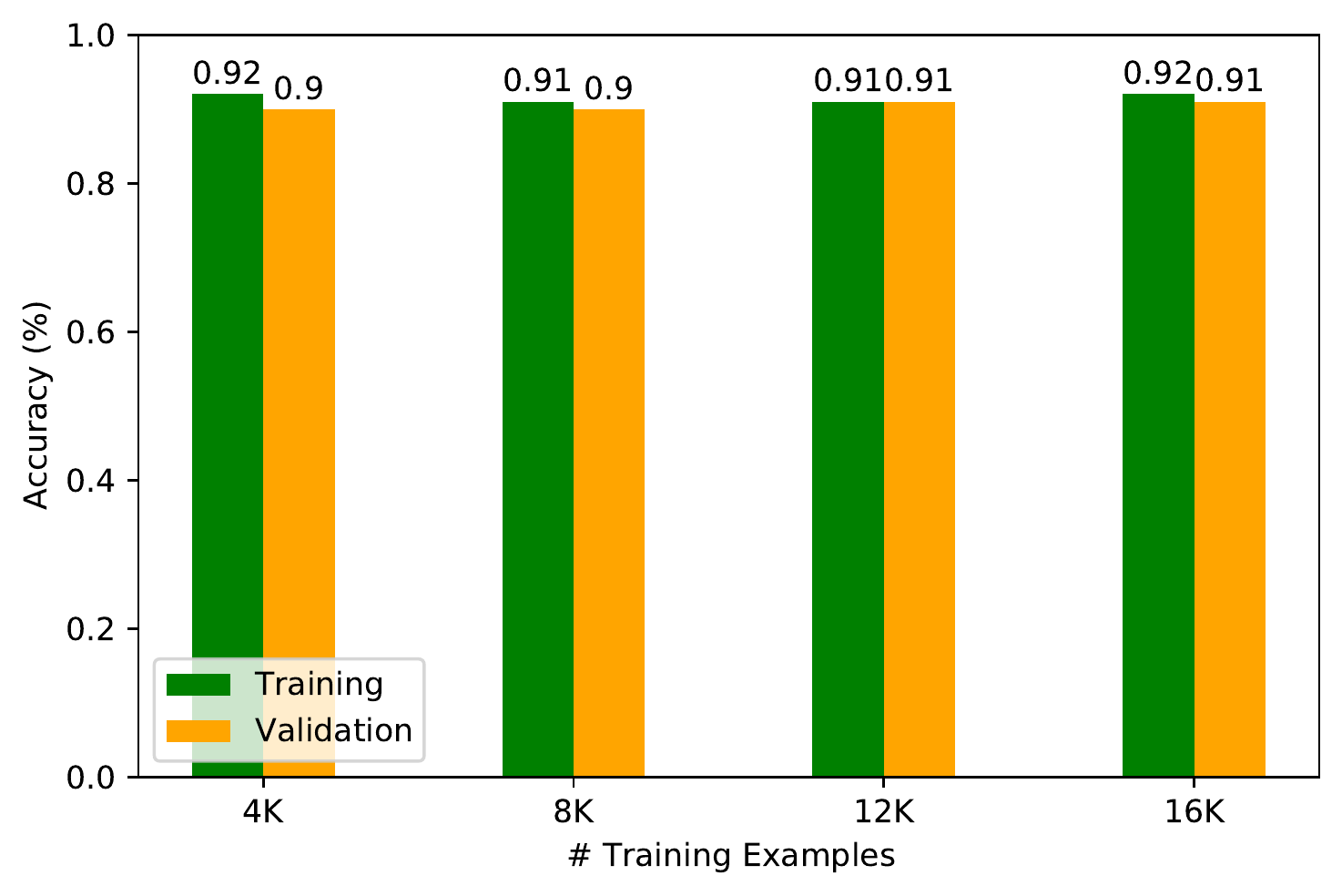}
 \caption{Custom-MalConv: IV3 progressive accuracy with CH.}
   \label{fig:Prog_CH}
\end{minipage}
%\hspace{-0.1cm}
\begin{minipage}[b]{0.5\linewidth}
 \centering
 \includegraphics[scale=.38]{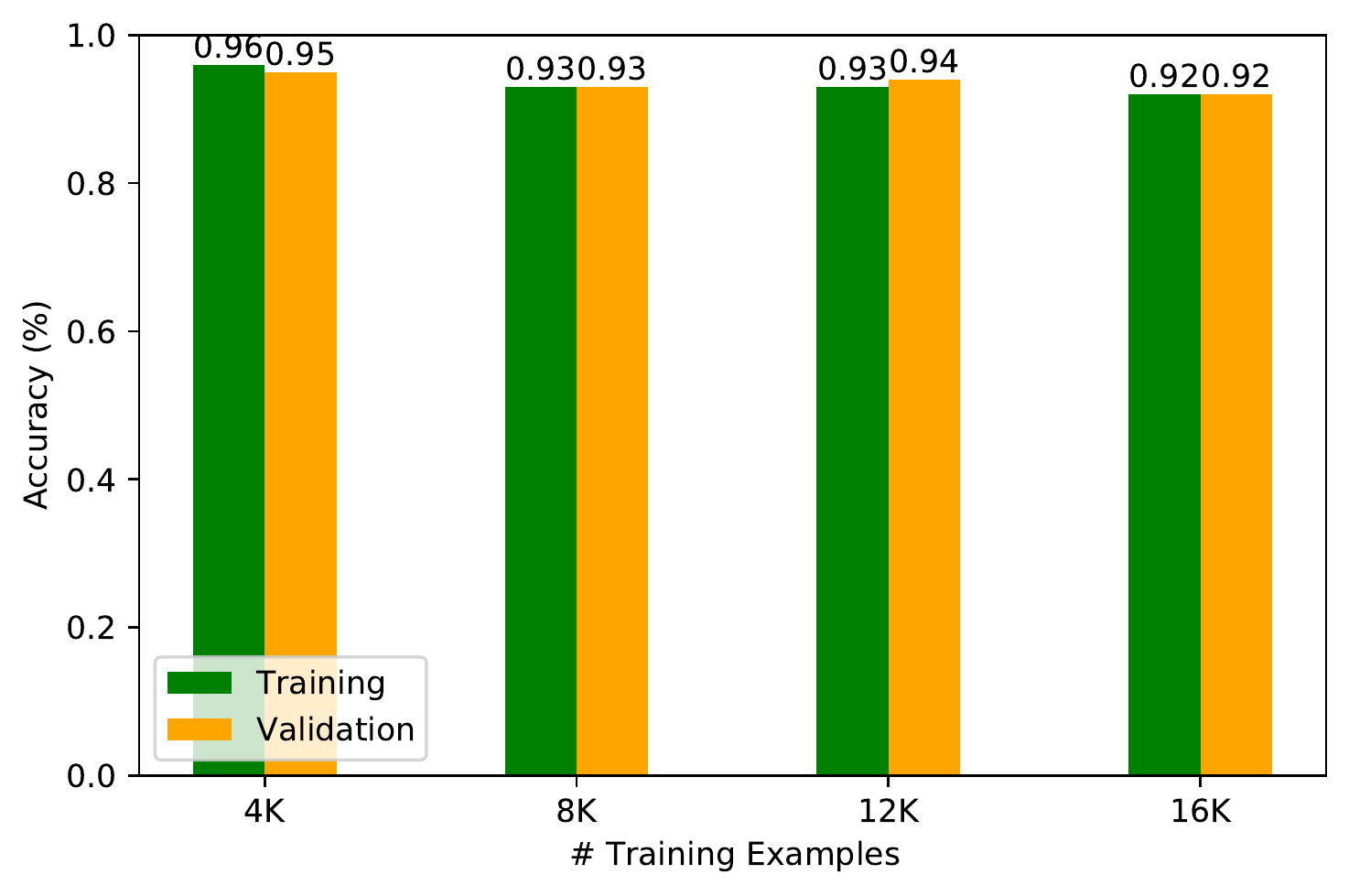}
 \caption{Custom-MalConv: IV3 progressive accuracy with EN.}
  \label{fig:Prog_EN}
\end{minipage}
\end{figure*}

\textbf{Custom-MalConv:}
Figure \ref{fig:Prog_CH} shows the training accuracy and validation accuracy of each progressive step in training $f_{s}$: IV3 with CH representation. It can be seen that the difference between training and validation accuracy narrows as the model progresses with more training data, which is an indication of the real-life effectiveness of the model on training examples it has never seen during training. Similarly, Figure \ref{fig:Prog_EN} shows the progressive training of $f_{s}$: IV3 with EN representation. Here, the trend persists, where we see a very high training accuracy across the board, but as we train on bigger data sets we get a more accurate model (validation accuracy is close to training accuracy).  

 For both CH and EN representations, it {\em takes  only 10\% the size of $f_{b}$'s training set to approximate it with $f_{s}$} that achieves 0.9 and 0.95 validation accuracy, respectively. On IV3 with CH, quadrupling the substitute training set (4K to 16K) improved its validation accuracy by 0.1 only, showing the {\em feasibility of approximation under data scarcity by taking advantage of transferability}. IV3 with EN shows a slightly different progress. On 16K approximation samples, it achieves 0.92 validation accuracy, which is 3\% less than the IV3 with CH trained on 4K samples. Looking at the difference between the training and validation accuracies (green vs. orange bars in Figures \ref{fig:Prog_CH} and \ref{fig:Prog_EN}), we notice that the narrower the gap between training and validation accuracy, the more stable a model would be classifying unknown samples.

\begin{figure*}[h!]

\begin{minipage}[b]{0.5\linewidth}
 \centering
 \includegraphics[scale=.41]{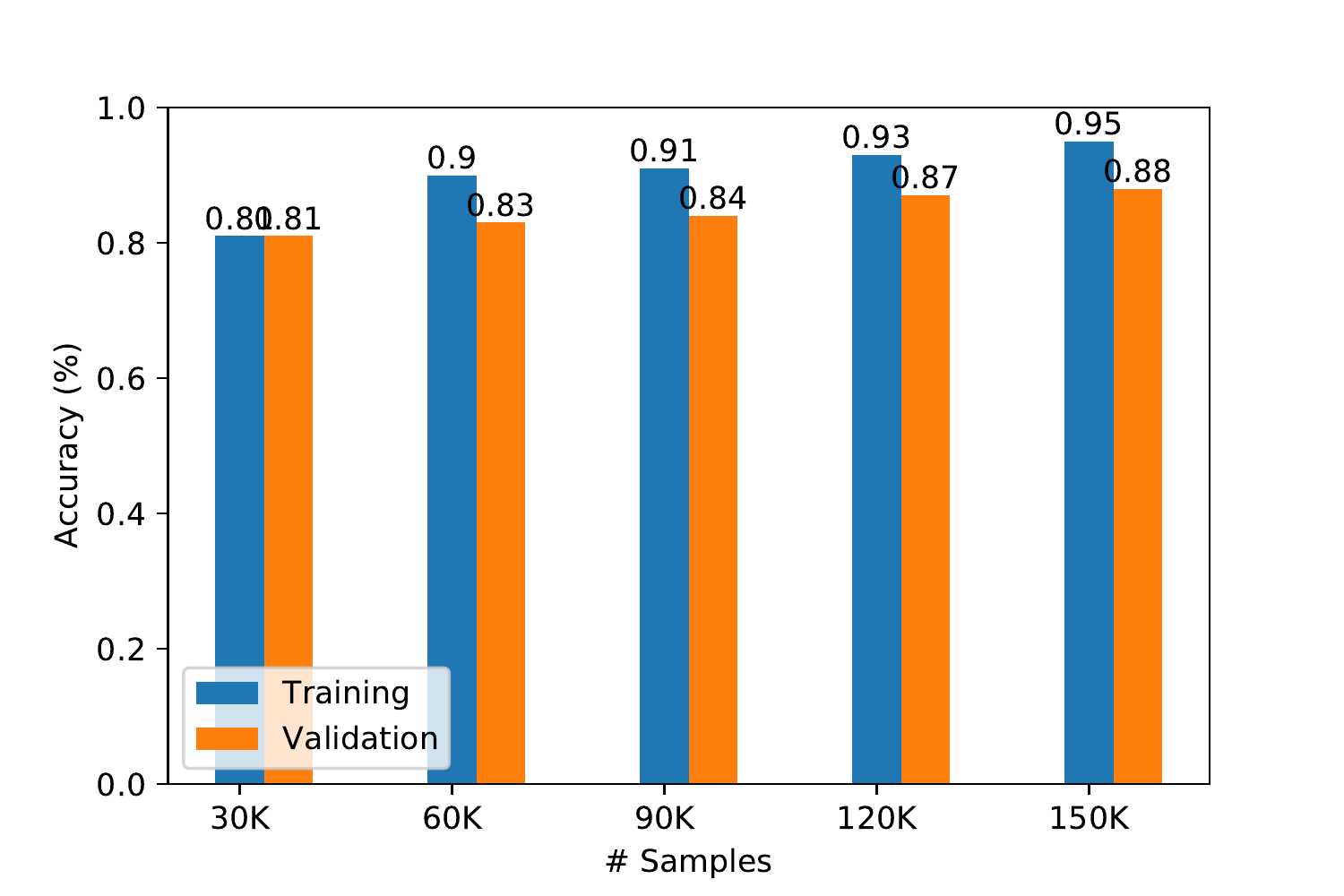}
  \caption{Progressive accuracy of $f_{s}$: LGBM.}
    \label{fig:LGBM-Progress}
\end{minipage}
%\hspace{0.1cm}
\begin{minipage}[b]{0.5\linewidth}
 \centering
 \includegraphics[scale=.41]{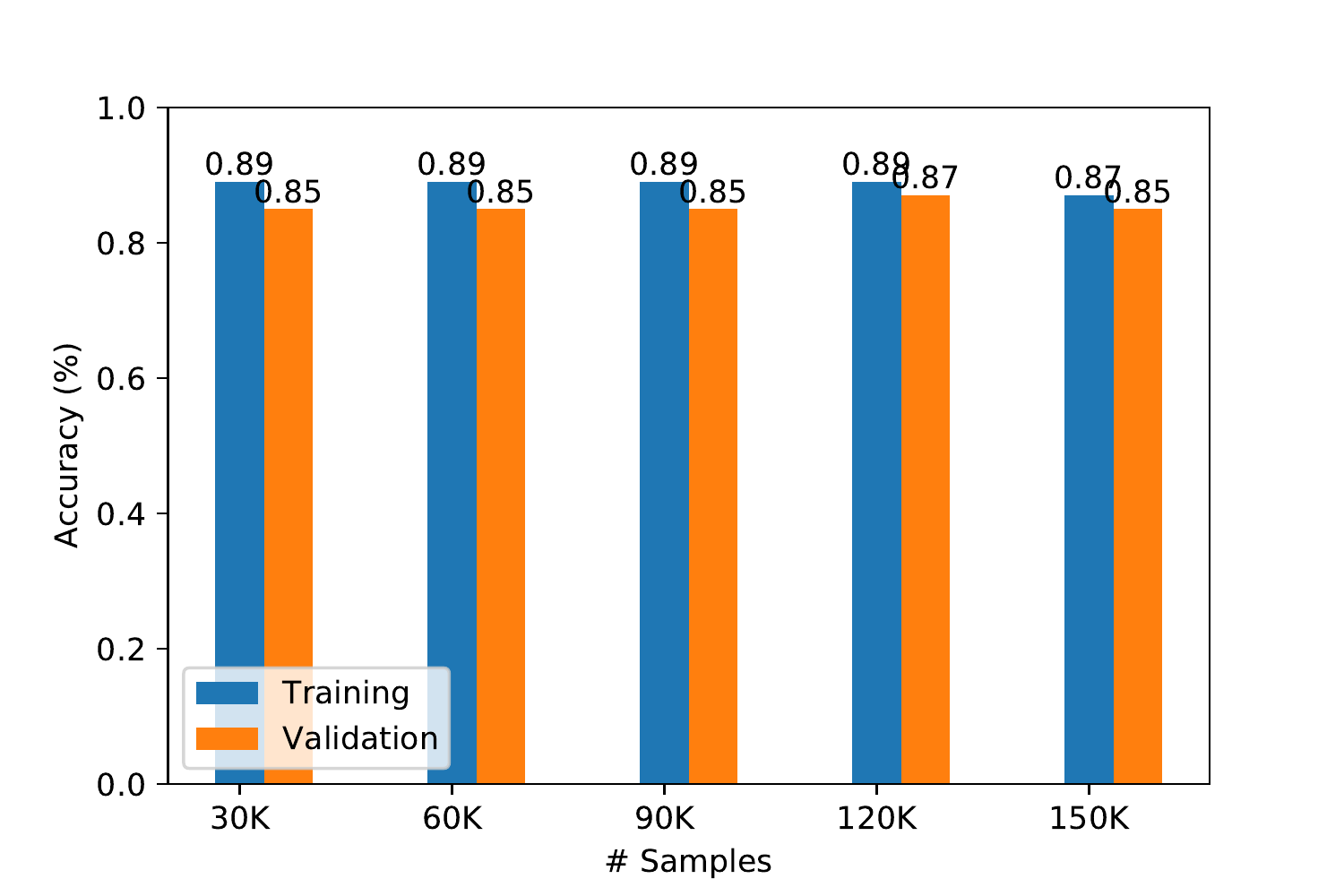}
 \caption{Progressive accuracy of $f_{s}$: DT.}
    \label{fig:DT-Progress}
\end{minipage}
\begin{minipage}[b]{0.5\linewidth}
 \centering
 \includegraphics[scale=.41]{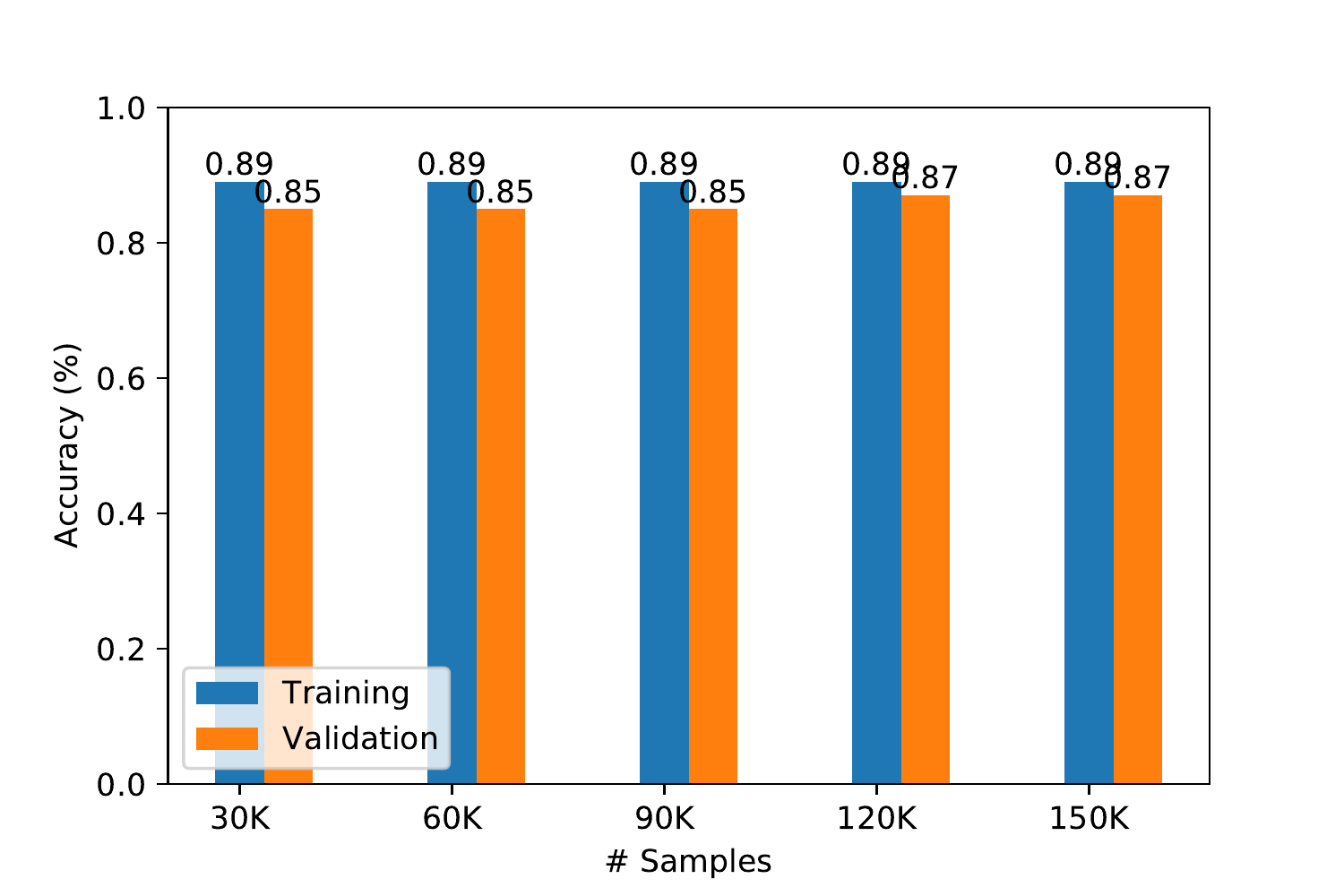}
 \caption{Progressive accuracy of $f_{s}$: RF.}
    \label{fig:RF-Progress}
\end{minipage}
%\hspace{0.1cm}
\begin{minipage}[b]{0.5\linewidth}
 \centering
 \includegraphics[scale=.41]{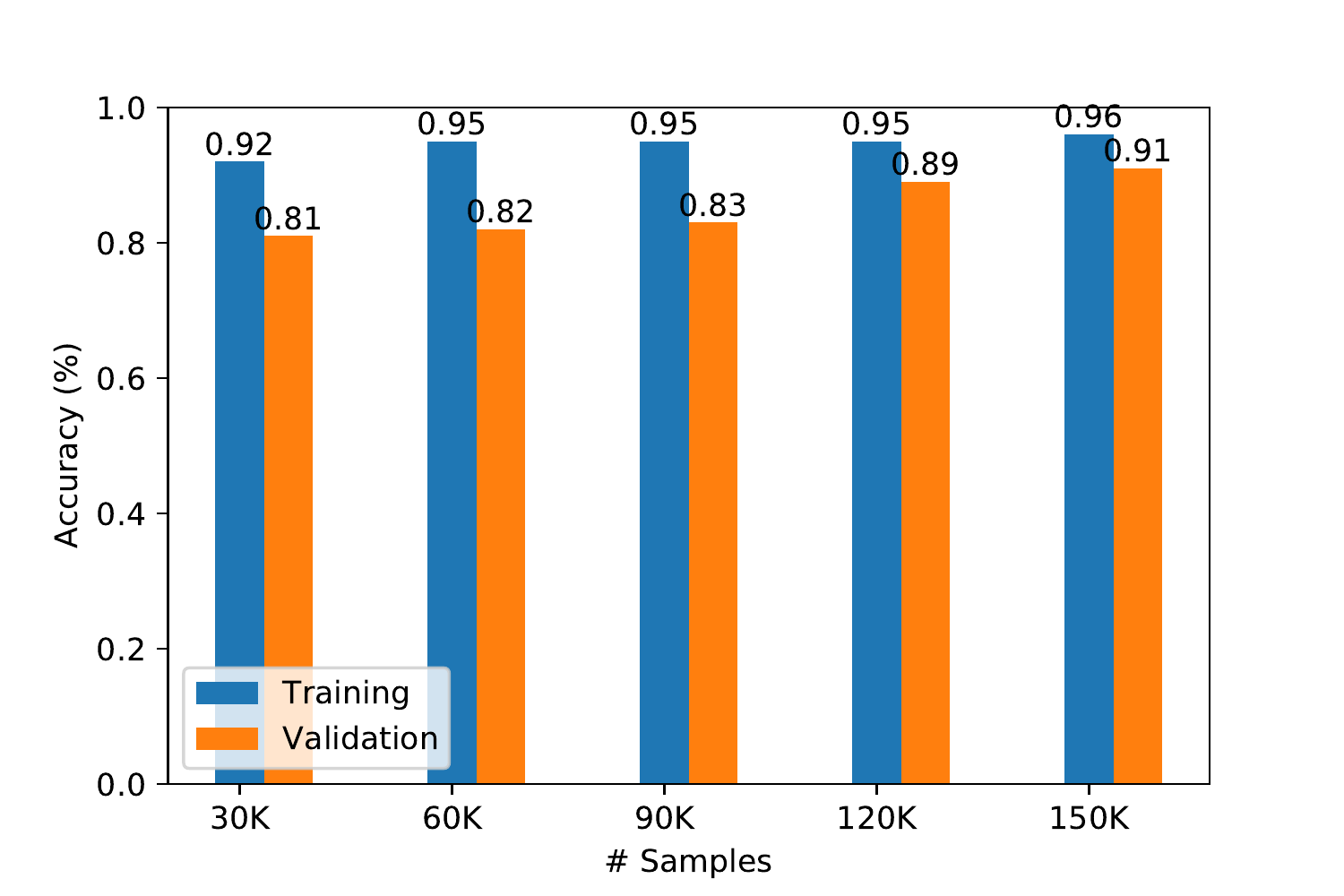}
 \caption{Progressive accuracy of $f_{s}$: $k$NN.}
    \label{fig:$k$NN-Progress}
\end{minipage}
\end{figure*}

\textbf{EMBER:} Figure ~\ref{fig:LGBM-Approx-Progress} shows the progressive validation accuracy of the four candidate substitute models against the LGBM black-box. Compared to the progress of Custom-MalConv substitutes (Figures \ref{fig:Prog_CH} and \ref{fig:Prog_EN}), the validation accuracies of LGBM, DT, and RF are relatively lower (in the range 0.85 - 0.88). However, $k$NN stands out as a clear contender (validation accuracy = 0.91) when compared with Custom-MalConv substitutes. Again, interestingly, the LGBM substitute, despite its architectural similarity to the target black-box LGBM, is not the best in terms of accuracy. From Figures \ref{fig:LGBM-Progress} - \ref{fig:$k$NN-Progress}, it is noteworthy that the gap between training and validation accuracy for EMBER substitutes is comparatively bigger (on average 4\%) compared to the narrow differences (on average 0.5\%) for Custom-MalConv substitutes shown in Figures \ref{fig:Prog_CH} and \ref{fig:Prog_EN}.

Looking at the substitute training set size with respect to the LGBM black-box, {\em to obtain the best performing substitute ($k$NN), it takes only 18.8\% (150K) of LGBM's training set size (800K)}. Interestingly, from Figure \ref{fig:LGBM-Progress}, we notice that the LGBM substitute, despite its architectural similarity with the target LGBM, performs relatively poorly, suggesting that {\em model architecture similarity may not always result in the best approximation}.

\subsection{Similarity Comparison Results}\label{subsec: similarity-results}

\textbf{Custom-MalConv:} As shown in Table \ref{tab:Results-malconv}, we have 3 substitutes, a Custom-MalConv substitute, InceptionV3 with CH, and InceptionV3 with EN. The comparison of these substitutes with the black-box is done on a separate comparison set, disjoint with the black-box training set and the substitute training set. 

\begin{table}[h!]
%\scriptsize
    \centering
\begin{tabular}{|l|r|r|}
\hline \textbf{Substitute ($f_{s}$)} & \textbf{Validation Accuracy} &\textbf{Similarity ($f_{b}$, $f_{s}$)}\\ \hline
    InceptionV3-ColorHilbert &  0.91& 82.19\% \\  \hline
    InceptionV3-Entropy & 0.92 & 88.65\%\\ \hline 
   Custom-MalConv-ByteSequence & 0.90& 80.11\%\\ \hline
   \end{tabular}
\vspace{0.075in}

\caption{Custom-MalConv: Similarity comparison between $f_{b}$ and $f_{s}$.}
\label{tab:Results-malconv}
\end{table}
On average, our approach achieved 83.7\% similarity score, with the highest similarity score of 89\%, on IV3 substitute trained on EN representation. The MalConv substitute that matches the architecture of the black-box Custom-MalConv model is the least accurate, which interestingly indicates that {\em model architecture similarity may not always result in a substitute that agrees well with a black-box target}. When we compare CH-based and EN-based substitutes, EN-based substitute outperforms the CH-based substitute by about 6.5\%, which could be attributed to the canvas coloring schemes of CH and EN. 

\begin{figure}[t!]
    \centering
    \includegraphics[scale=.56]{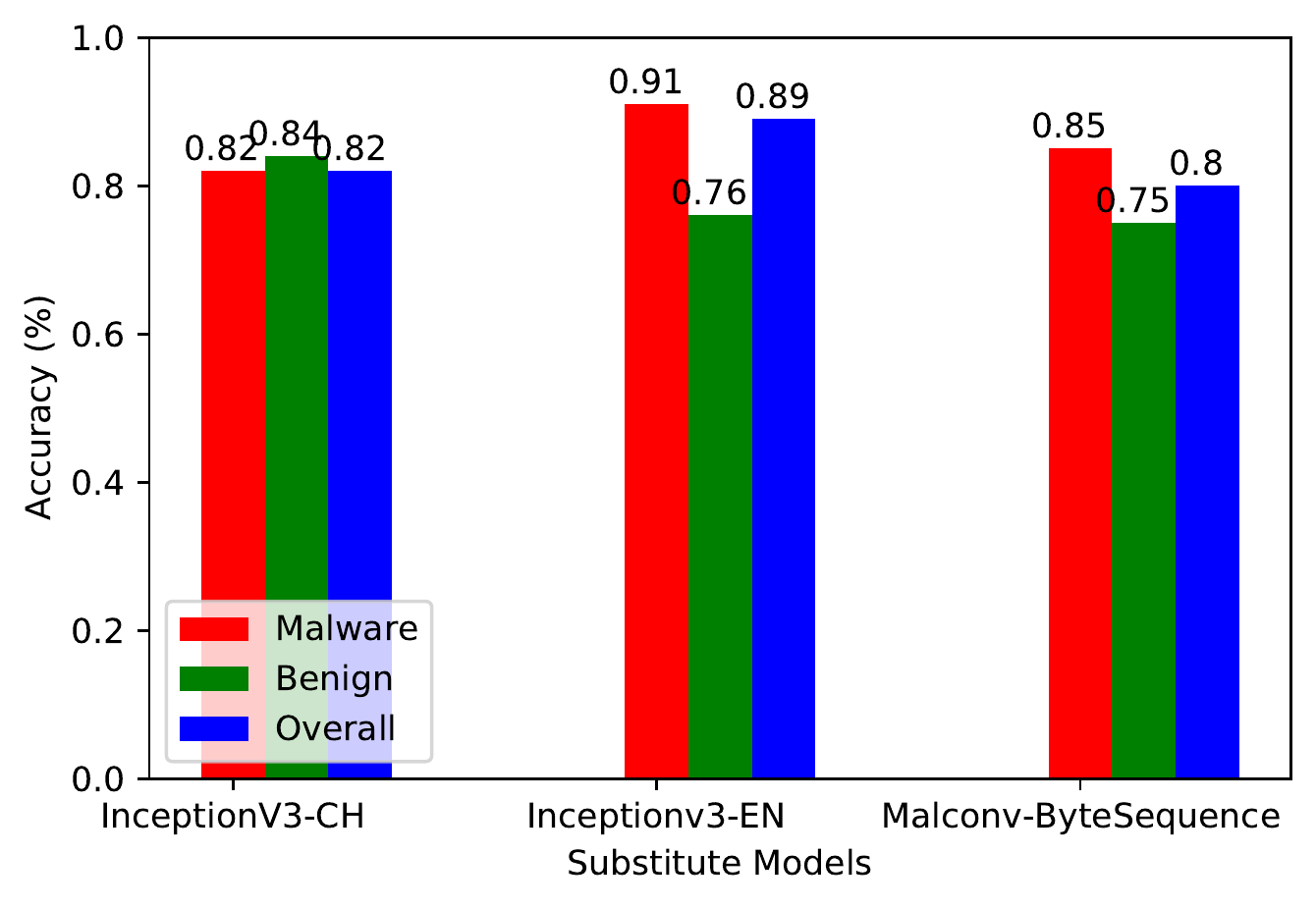}
    \caption{Custom-MalConv: Benign vs. malware prediction agreement split for $f_{b}$ and $f_{s}$.}
    \label{fig:malconv-split}
\end{figure}

Figure \ref{fig:malconv-split} shows the details of similarity scores for the three substitute models split into malware and benign. To compute the split, for each label, we divide the number of agreements between $f_{b}$ and $f_{s}$ by the total number of samples for that label. Having a malware detection rate higher than the benign detection rate serves us best, since our goal is to be able to identify malware identified as so by the black-box. It can be seen that the EN-based substitute tends to agree with the black-box more on malware, while the CH-based substitute agrees with the black-box more on the benign predictions. Again, this variation is rooted in the canvas coloring methods of the two representations discussed earlier.

\begin{figure*}[t!]
\begin{minipage}[b]{0.5\linewidth}
 \centering
 \includegraphics[scale=.38]{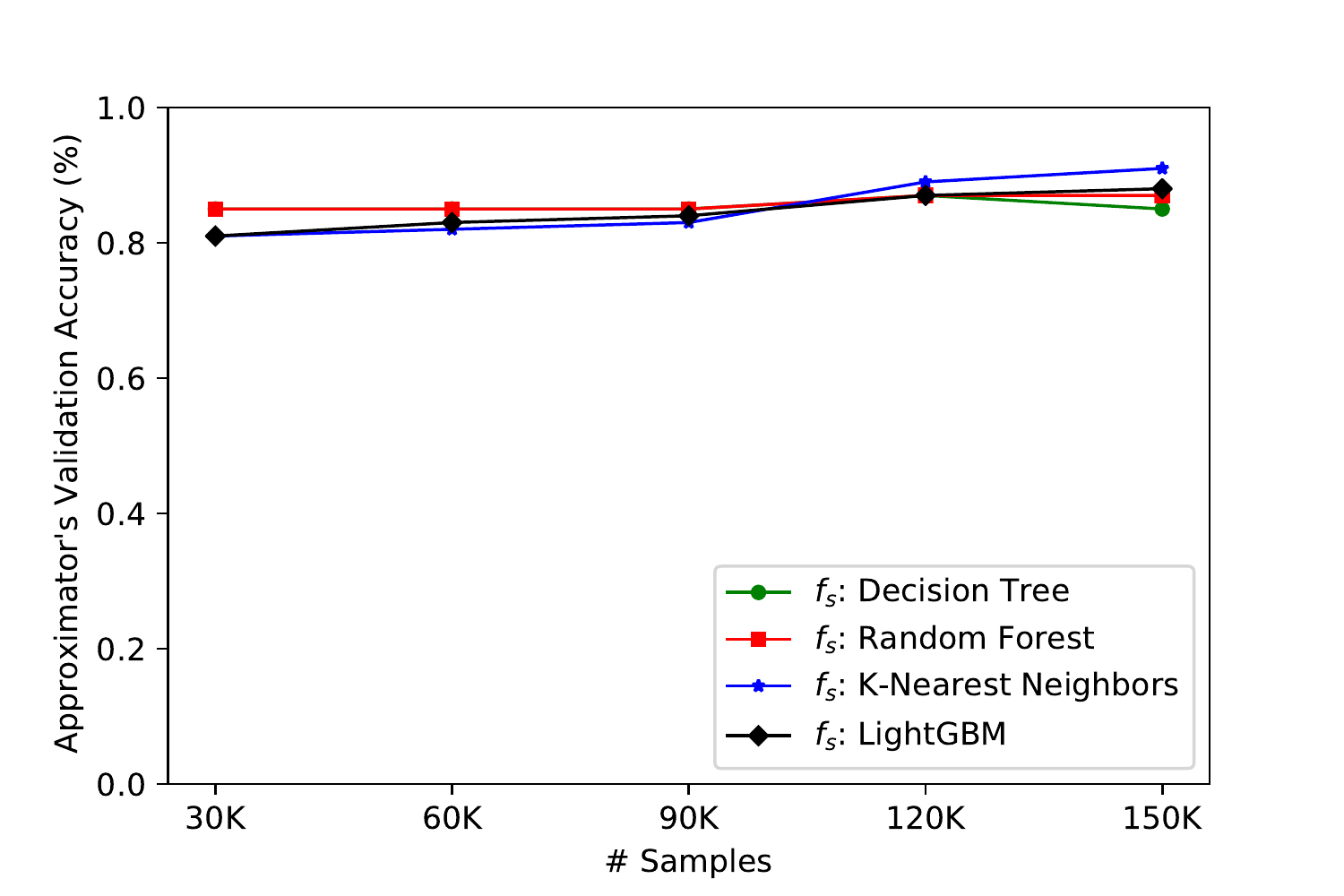}
  \caption{EMBER: progressive validation accuracy of candidates for $f_s$.}
    \label{fig:LGBM-Approx-Progress}
\end{minipage}
%\hspace{0.1cm}
\begin{minipage}[b]{0.5\linewidth}
 \centering
 \includegraphics[scale=.38]{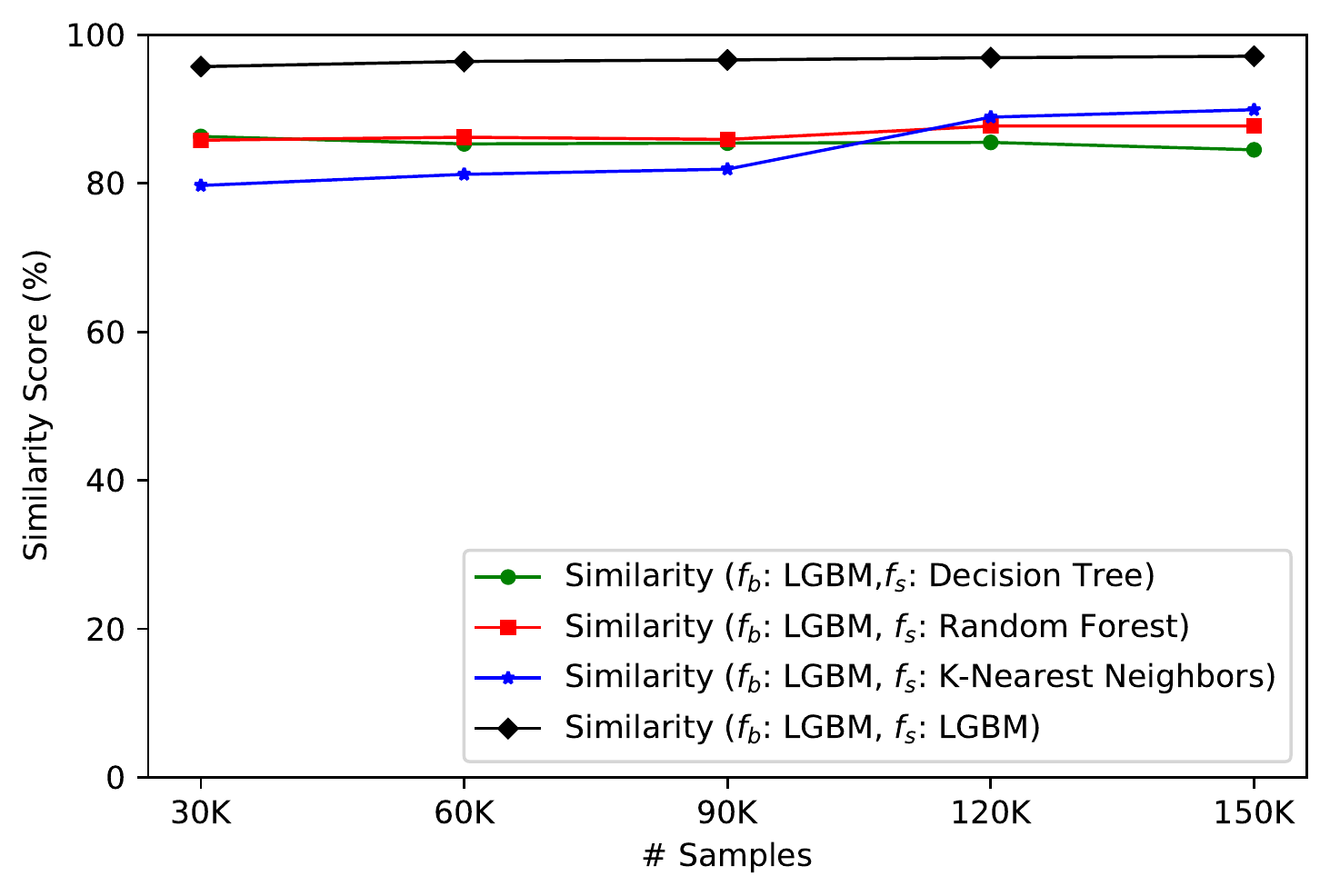}
\caption{EMBER: progressive similarity of $f_{b}$ with $f_{s}$.}
 \label{fig:LGBM-Sim-Progress}

\end{minipage}
\end{figure*}

\textbf{EMBER:}
Table \ref{tab:Results-ember} summarizes the similarity comparison between LGBM black-box and candidate substitute models. On average a substitute model agrees with the black-box on 90.3\% of the comparison samples. $k$NN, as the closest approximation with a different model architecture from the black-box,  agrees with it on almost 90\% of the comparison samples. This result is comparable to the best substitute in Custom-MalConv: InceptionV3-Entropy with 89\% similarity score. We also notice from Table \ref{tab:Results-ember} that, although it is the least accurate model from Figure \ref{fig:LGBM-Approx-Progress}, LGBM substitute agrees with the LGBM black-box on over $97$\% of the comparison instances. This is not surprising, given the exactness of the model architectures. It is, however, in contrast to our observation on the similarity of Custom-MalConv-ByteSequence substitute with the same black-box model architecture, which scores just over 80\% similarity (see Table \ref{tab:Results-malconv}).

\begin{table}[t!]
 %\scriptsize
    \centering
\begin{tabular}{|l|r|r|r|}
\hline \textbf{Substitute ($f_{s}$)} & \textbf{Validation Accuracy} &\textbf{Similarity ($f_{b}$, $f_{s}$)}\\ \hline
    Decision Tree  &0.85 & 86.3\% \\ \hline
    Random Forest  &0.87 & 87.7\%\\ \hline
    $k$-Nearest Neighbors  & 0.91& 89.9\%  \\ \hline 
   LightGBM  & 0.88 & 97.1\%  \\ \hline
    \end{tabular}
\caption{EMBER: Similarity comparison between $f_{b}$ and  $f_{s}$.}
\label{tab:Results-ember}
\end{table}
%\vspace{-.25in}

Figure \ref{fig:LGBM-Sim-Progress} shows the evolution of similarity scores showing how $k$NN substitute's similarity improves almost by 10\% in the range 90K - 150K of approximation instances. RF substitute also shows a similar pattern, but with a smaller improvement of 4\% (83\% to 87\%). Note that the LGBM substitute remained stable across the progressive increment of the substitute training set.

%\vspace{-.25in}

\begin{figure*}[h!]
\begin{minipage}[b]{0.5\linewidth}
 \centering
 \includegraphics[scale=.40]{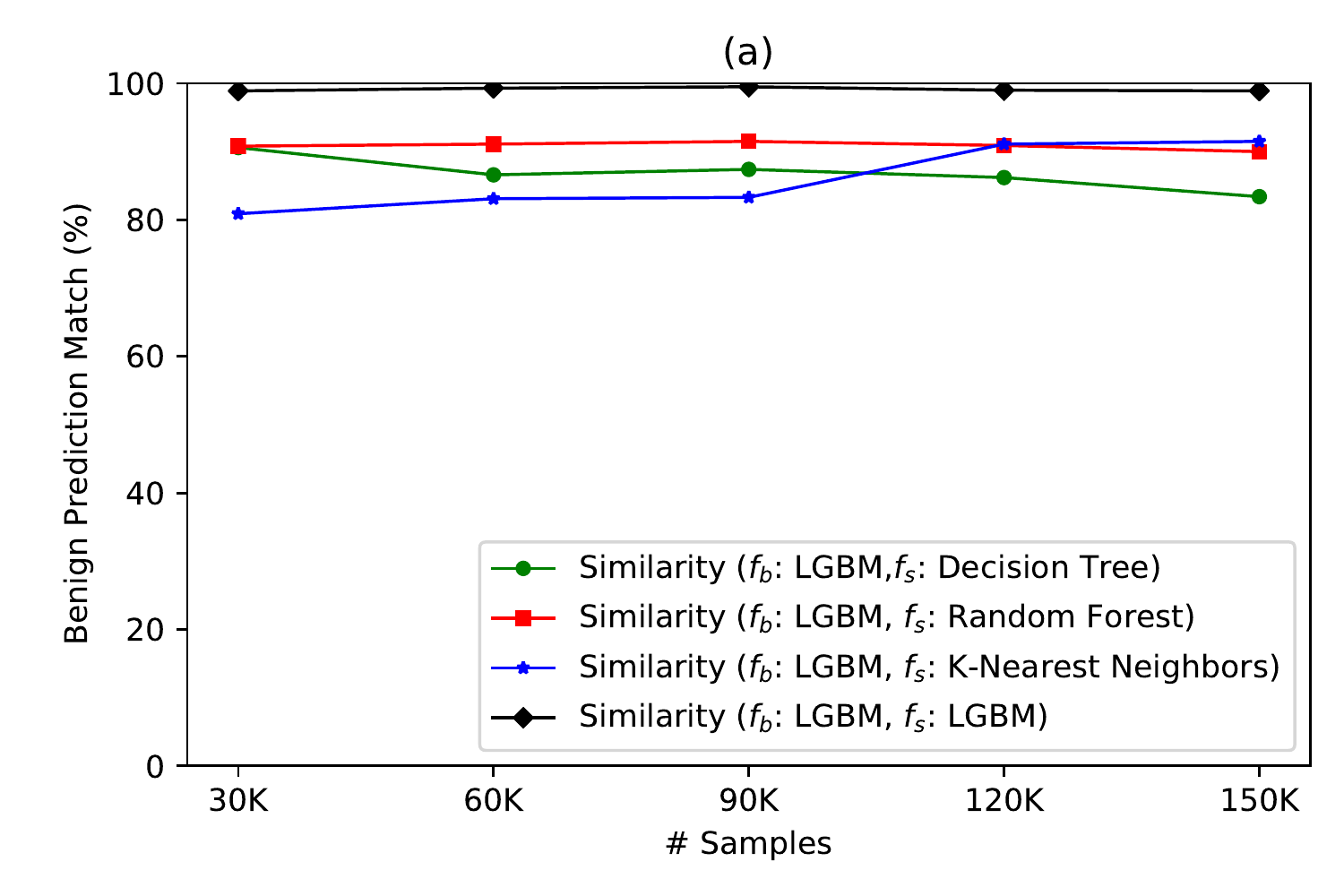}
  \caption{EMBER: Benign prediction match between $f_{b}$ and $f_{s}$.}
    \label{fig:EMBER-benign-split}
\end{minipage}
%\hspace{0.1cm}
\begin{minipage}[b]{0.5\linewidth}
 \centering
 \includegraphics[scale=.40]{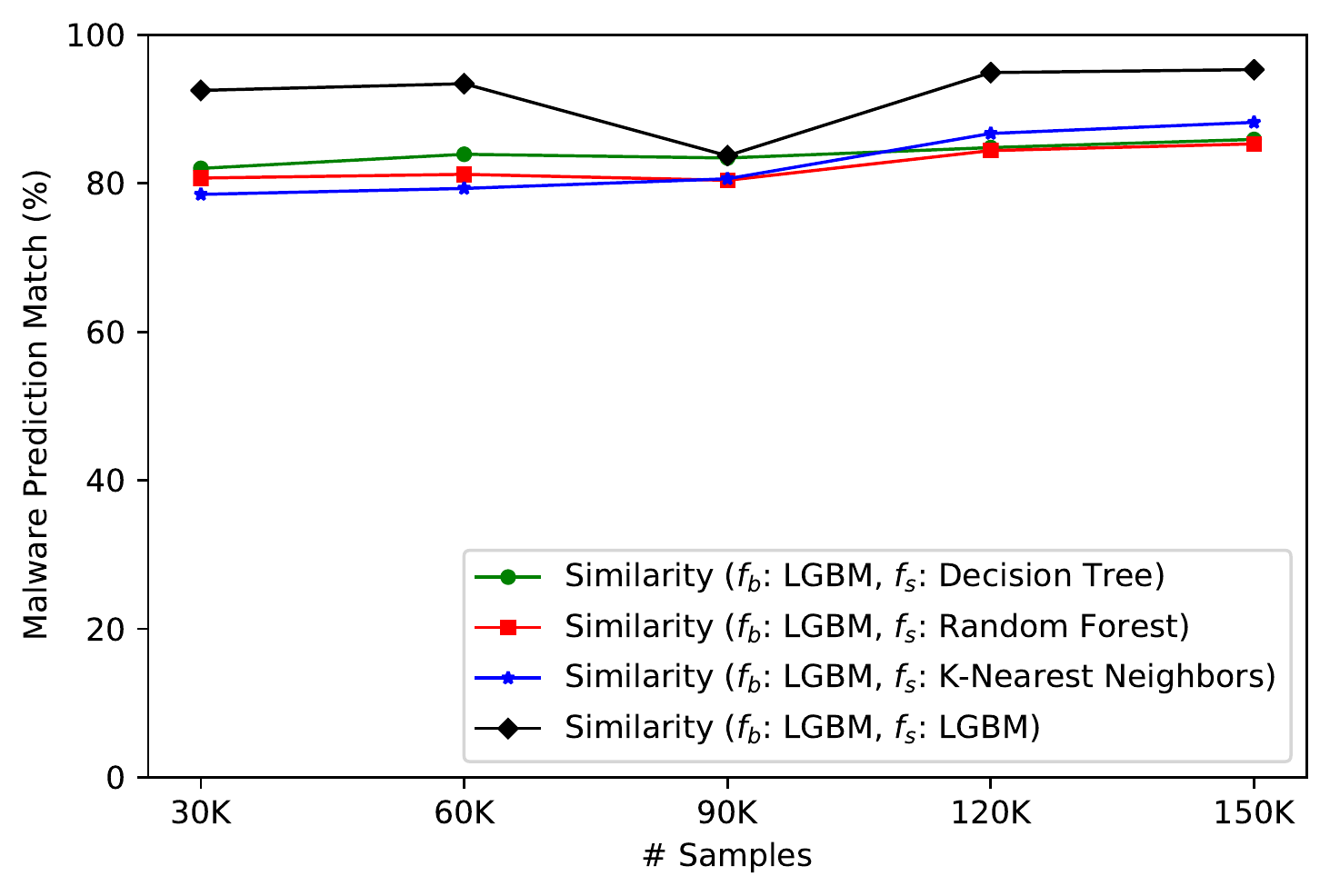}
	   \caption{EMBER: Malware prediction match between $f_{b}$ and $f_{s}$.}
	    \label{fig:EMBER-malware-split}

\end{minipage}
\end{figure*}
%\vspace{-.1\in}

Finally, Figures \ref{fig:EMBER-benign-split} and \ref{fig:EMBER-malware-split} show the progression of benign and malware similarity matches as we train the substitutes. It can be seen that, overall, there is high degree of agreement between the LGBM black-box and the substitutes on benign predictions. On benign predictions, the LGBM substitute shows a stable and close to 100\% agreement with its black-box counterpart. On malware predictions, it shows variations in the range 83\% - 95\%.
% !TEX root =  main.tex
\section{Related Work}\label{chap:related}
We review closely related work in image classification and malware detection, and make approach-level comparisons in Table~\ref{tab: related-comparison}.

Tramer  et al.~\cite{model-stealing16} show equation-solving and optimization based attacks that approximate target ML models with near-perfect replication for popular learning algorithms including logistic regression, support vector machines, neural networks, and decision trees. They demonstrate their attacks against production MLaaS providers such as BigML and Amazon ML. They note that their most successful model approximations are for MLaaS services that return prediction labels with probability scores.  

Papernot et al.~\cite{PracticalBBAtk} demonstrate
an adversary controlling a remotely
hosted DNN with no knowledge about the DNN and its training data. They train a local substitute against a target DNN using synthetically generated inputs labeled by the target DNN. They then use the local substitute to craft adversarial examples to evade DNNs hosted by MetaMind, and also train logistic regression substitutes against Amazon and Google prediction APIs.

Papernot et al.~\cite{transferability16} train substitute models to approximate a target black-box model, and craft adversarial examples to evade it. They use the {\em Jacobian-based dataset augmentation} method to synthetically generate examples for training the substitute.  Similar to \cite{PracticalBBAtk}, they demonstrate transferability within and between different classes of ML models such as DNNs, logistic regression, support vector machines, decision trees, nearest neighbors, and ensembles.

Hu and Tan~\cite{MalGAN17} train a substitute model using the GAN framework to fit a black-box malware detector. In a follow-up work \cite{RNN-Black-Box18}, they use the GAN-based substitute model training for a recurrent neural network. Both approaches (\cite{MalGAN17,RNN-Black-Box18}) assume the adversary knows the type of features and the architecture of the black-box model.
In our case, we assume that the adversary knows nothing about the details of the black-box. In addition, our approach differs from \cite{MalGAN17} in the assumption about the underlying feature representations of training examples (Windows PEs). 

While \cite{MalGAN17} assumes the same (API calls precisely) feature representation of the black-box and the substitute, in our work, the black-box and the substitute have different feature representations.
Rosenberg et al. \cite{APIBlackBoxEvade18} adopt the Jacobian-based augmentation \cite{transferability16} to synthesize training examples to approximate and evade a target black-box malware detector based on API call features.
Like Papernot et al. \cite{transferability16} and Hu and Tan \cite{MalGAN17}, Papernot  et al. in a different work \cite{PracticalBBAtk} craft adversarial examples to evade a black-box model. 

Orekondy et al.~\cite{KnockoffNets19} follow a similar threat model to ours where publicly available images from a domain different from the black-box model are used to train a substitute `Knockoff' model. They show selecting samples from a totally different distribution using reinforcement learning is effective for an approximation attack, while using different model architectures for the victim black-box and the substitute knockoff. We note that, in the cross-domain transferability setting, this work is the closest to ours except that it uses the same pixel features for both $f_b$ and $f_s$, while we use byte sequences for $f_b$ and pixels for $f_s$.

Next, we make approach-level comparisons of our work and closely related prior work. Table \ref{tab: related-comparison} shows multi-criteria comparison of this work with the state-of-the-art in image classification and malware detection. Since direct quantitative comparison is non-trivial given differences in assumptions and dataset/task heterogeneity, our comparison is rather qualitative and is based on assumptions about feature representation, model architecture, initial substitute training set size, training dataset overlap between black-box and substitute model, and black-box prediction output.

\textbf{Features ($f_{b}$, $f_{s}$)} compares assumptions about features of the black-box ($f_{b}$) and the substitute ($f_{s}$). While prior work \cite{transferability16,MalGAN17,EfficientStealing19,RNN-Black-Box18,PracticalBBAtk,APIBlackBoxEvade18,PRADA19} assumed similar features for $f_{b}$ and $f_{s}$, we consider raw-bytes for $f_{b}$ and pixels for $f_{s}$.

\textbf{Model ($f_{b}$, $f_{s}$)} captures assumptions about similarity of model architectures for $f_{b}$ and $f_{s}$. Unlike most prior work \cite{MalGAN17,RNN-Black-Box18,transferability16,PracticalBBAtk,PRADA19,EfficientStealing19} which assume the same model architecture for $f_{b}$ and $f_{s}$, we assume the adversary is free to evaluate different model architectures for $f_{s}$ (hence could match $f_{b}$'s architecture or end up using a different one). 
\begin{table*}[!htbp]
\scriptsize
\centering
\begin{tabular}{|c*{8}{|c}|}\hline
%\backslashbox{}{}
&\makebox[3em]{\textbf{\cite{PracticalBBAtk,PRADA19}}}
&\makebox[3em]{\textbf{\cite{model-stealing16}}}&\makebox[3em]{\textbf{\cite{APIBlackBoxEvade18}}}&\makebox[3em]{\textbf{\cite{EfficientStealing19}}}&\makebox[3em]{\textbf{\cite{MalGAN17,RNN-Black-Box18}}}&\makebox[3em]{\textbf{\cite{KnockoffNets19}}}&\makebox[3em]{\textbf{\cite{transferability16}}}&\makebox[5em]{\textbf{This work}}\\\hline
\textbf{Features($f_{b}$, $f_{s}$)}      &same   &same  &same  & same &same &same   &same   &different\\ \hline
\textbf{Model($f_{b}$, $f_{s}$)}     &same   &same/different  &different   &same &same &different   &different   &different\\ \hline
\textbf{Seed-set($f_{s}$)}   &moderate   &moderate   &moderate    &limited &moderate &moderate  &limited    &limited\\ \hline
\textbf{Data($f_{b}$, $f_{s}$)}      &disjoint    &disjoint    &N/A &N/A &disjoint &disjoint    &disjoint   &disjoint\\ \hline
\textbf{Output ($f_{b}$)} &label+conf.   &label+conf.   &label   &label &label &label   &label  &label \\ \hline 
\end{tabular}
\caption{Approach-level qualitative comparison with closely related work.}
\label{tab: related-comparison}
\end{table*}
\vspace{-.1in}

\textbf{Seed-set ($f_{s}$)} compares assumptions on the size (number of training examples) of the seed-set to train $f_{s}$. In \cite{transferability16,PracticalBBAtk,EfficientStealing19}, the adversary explores synthetic data generation techniques such as augmentation to generate more training samples to train $f_{s}$. In \cite{MalGAN17} and \cite{RNN-Black-Box18}, the adversary collects enough samples to train $f_{s}$. In this work, we explore approximation with access to a limited seed-set that we extend with data augmentation techniques.

\textbf{Data ($f_{b}$, $f_{s}$)} examines whether or not there is overlap in training sets used for $f_{b}$, $f_{s}$, and comparison of $f_{b}$ with $f_{s}$. In this work, we use disjoint datasets for training $f_{b}$, $f_{s}$, and comparison of $f_{b}$ with $f_{s}$. Doing so enables assessment of the effectiveness of $f_{s}$ approximated with a completely new dataset.

\textbf{Output ($f_{b}$)} captures whether label only or label with confidence score (conf.) is returned from $f_{b}$. While prior work used label only \cite{MalGAN17,APIBlackBoxEvade18,EfficientStealing19,RNN-Black-Box18,KnockoffNets19} and label with probability score \cite{PracticalBBAtk,model-stealing16,PRADA19}, our black-box returns only the label of the PEs, i.e., ``benign'' or ``malware''. 
% !TEX root =  main.tex
\section{Conclusion}   \label{chap:concl}

We presented a best-effort adversarial approximation approach that leverages representation mapping and cross-domain transferability to obtain a close-enough approximation of a black-box malware detector. We show that an adversary can obtain nearly 90\% similarity between a black-box model and its approximation beginning with a limited input-set to the black-box, different features and disjoint training sets for the black-box and its substitute. We further demonstrate that our approach generalizes to multiple model architectures across different feature representations. This work broadens the scope of adversarial approximation of a black-box ML model in a strictly black-box setting. Our results shade light on the fact that different feature representations may not necessarily hinder close-enough approximation and disjoint datasets still result in successful approximation. More importantly, a pre-trained multi-class image classifier, such as Inception V3, can be re-purposed to approximate a binary malware classifier, demonstrating the scope of transferability beyond cross-model and extending it to a cross-domain setting.

%\pagebreak
\bibliographystyle{splncs04}
\bibliography{main}
% !TEX root =  main.tex
\section*{Appendix A}\label{app-a} 
\textbf{Augmentation on Image Representation of PEs: }
Training a substitute based on a pre-trained model depends on the availability of enough training examples, but the adversary may not have the leverage to collect enough training data. This challenge is specially true when one attempts to collect benign PEs from the wild. We observe that, while it is relatively easy to obtain a dataset of hundreds of thousands of malware PEs from malware repositories, it takes weeks to collect benign executables due to the lack of publicly curated benign PEs. To address the scarcity of training examples, in other domains such as image classification, the adversary exploits the notion of {\em data augmentation}, which aims to synthesize minimally manipulated yet semantically intact variants of an image representation of the PEs \cite{transferability16,PracticalBBAtk}. Intuitively, augmentation involves slightly altering an image while keeping the main features of the image intact. In vision tasks (e.g., image classification), augmentation methods (e.g., slight rotation, flipping) shouldn't visually alter the object. In our approach, however,  augmentation needs to keep PE semantics (e.g., malicious payload) intact. While slightly altering a PE's image, we don't want to end up with a mutated image that semantically diverges from the original bytes representation of a PE.

\begin{figure}[]
    \centering
        \includegraphics[scale =.5]{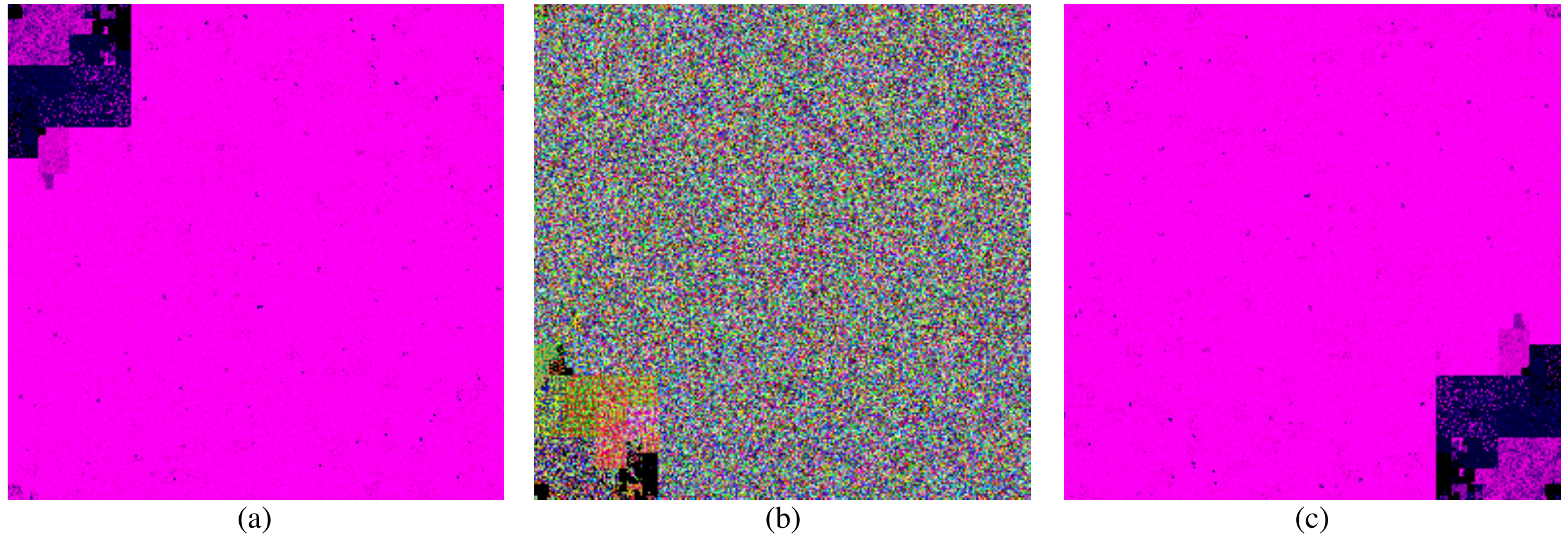}
        \caption{Augmentation of a benign PE (02Micro\_Card\_Reader\_Driver\_3.11.exe). (a): original EN, (b): flipped CH, (c): rotated EN.}
     \label{fig: Aug-exempls-Benign}
\end{figure} 

\begin{figure}[]
    \centering
        \includegraphics[scale=.5]{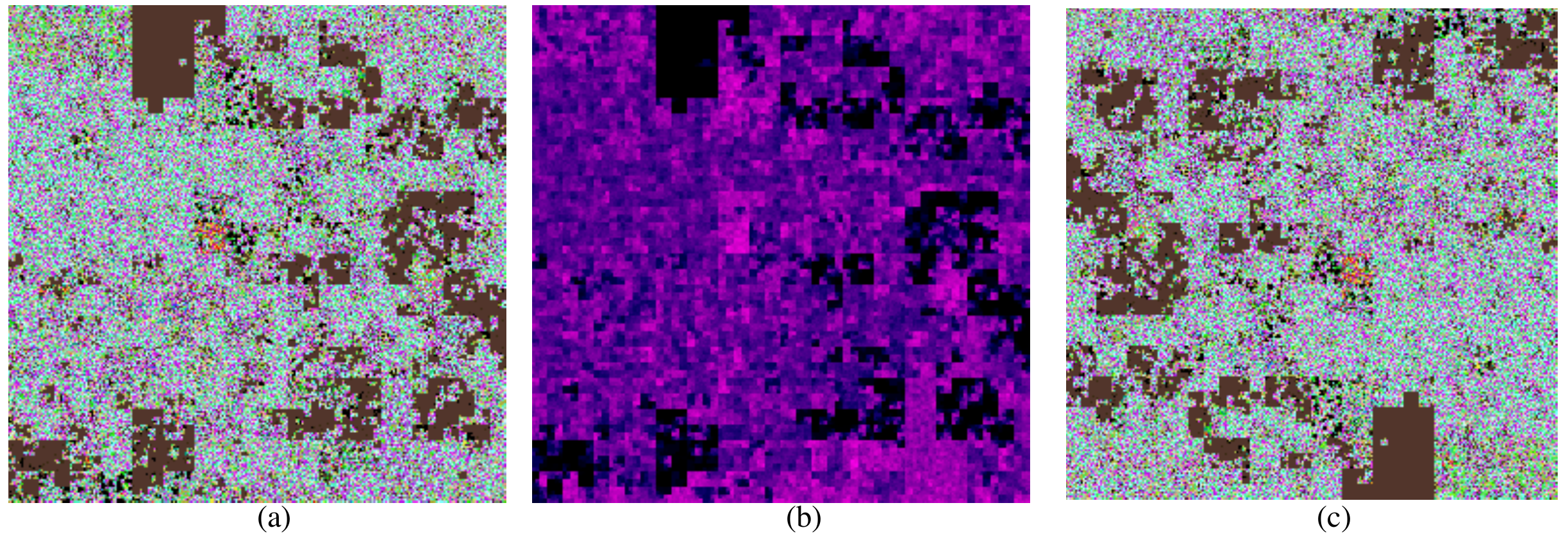}
        \caption{Augmentation of a malware PE (Trojan.GenericKDZ.58985). (a): original CH, (b): flipped EN, (c): rotated CH.}
     \label{fig: Aug-exempls-Malware}
\end{figure} 

\begin{figure}[htbp]
    \centering
    \includegraphics[scale=.55]{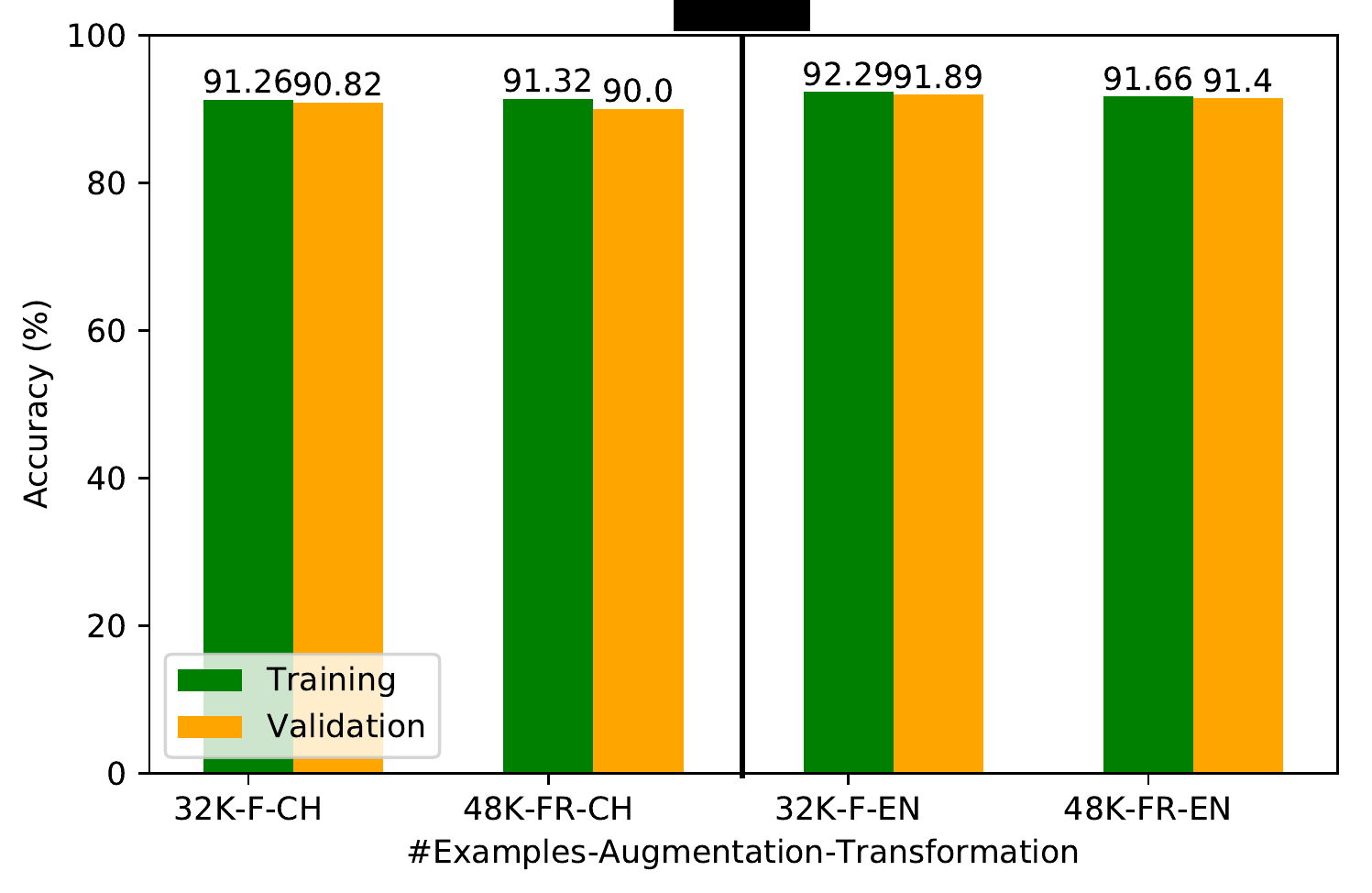}
    \caption{Custom-MalConv: accuracy of $f_{s}$ for Color Hilbert (CH) and Entropy (EN) representation with image flipping (F) augmentation  and rotation (R) augmentation.}
    \label{fig:Aug_CH}
\end{figure}

To explore the potential of augmentation, we carefully choose the augmentation methods for our approximation strategy. In particular, we selected {\em flipping} and {\em rotation} to synthesize variants of the training examples. These two augmentation methods are proved to keep the original structure of an image intact \cite{PracticalBBAtk}. Figures \ref{fig: Aug-exempls-Benign} and \ref{fig: Aug-exempls-Malware} illustrate flipping and rotation of a benign and malware sample, respectively.  Through augmentation, we managed to increase our substitute training set by three-fold. The impact, on the substitute's accuracy, of extending the initial substitute training set through augmentation is shown in Figure~\ref{fig:Aug_CH}.

\textbf{Augmentation-Based Approximation:}\label{subsec: aug-results}
The goal here is emulating an adversary with scarce training data and aims to expand to more training data with less time. We begin with the 8K executables for each label (16K overall) and apply $(a)$ rotation (by 90 degrees), $(b)$ flipping, and $(c)$ rotation + flipping to obtain twice as much (32K) for rotation and flipping each, and three times (48K) by merging the samples synthesized by rotation and flipping with the original 16K samples we started with. In Figure \ref{fig:Aug_CH}, the left-half shows substitute accuracy for CH and the right-half shows the same, but for EN. The CH results are not much different from the results from Figure \ref{fig:Prog_CH} and \ref{fig:Prog_EN}. This doesn't mean that the augmentation strategy is not useful, but it gives us visibility into the effectiveness of the specific representation and augmentation we used. 
For EN results, however, we see a slightly different trend. In particular, we notice a small improvement when using the flipped images with our original set. This could be attributed to the different features in the EN representation that the model is trained on. However, we see a very minor drop in accuracy (91.8\% $\rightarrow$ 91.4\%) when we add the rotated images. The accuracy of augmentation methods is overall comparable to the progressive training presented earlier. More importantly, so long at PE semantics is preserved, the augmentation methods can be done fast to address the data scarcity challenge for the adversary. 

\textbf{Similarity Comparison with Augmentation:} The last 4 rows in Table~\ref{tab:Results-aug} show substitutes trained with augmented data for both EN and CH representations. Comparing the different augmentation methods, it depends on whether we use CH or EN. The EN-based representations consistently outperformed the CH-based representations for all the augmentation techniques. As discussed earlier, this difference is attributable to the different canvas coloring  methods used in CH and EN.

\begin{table}[h!]
%\scriptsize
    \centering
\begin{tabular}{|l|r|r|r|}
\hline \textbf{Approximation Method} & \textbf{Similarity ($f_{b}$, $f_{s}$)}\\ \hline
    Custom-MalConv-ByteSequence & 80.11\\
    InceptionV3-ColorHilbert & 82.19 \\ 
    InceptionV3-Entropy & 88.65\\ 
    \hline 
    InceptionV3-ColorHilbert(32K) Flipped  & 81.69  \\ 
    InceptionV3-Entropy(32K) Flipped  & 88.15  \\
    InceptionV3-ColorHilbert (48K) Flipped and Rotated  & 80.76  \\ 
    InceptionV3-Entropy (48K) Flipped and Rotated  & 88.88  \\ \hline
\end{tabular}

\caption{Custom-MalConv: similarity comparison including the augmentation-based substitutes.}
\label{tab:Results-aug}
\end{table}

\end{document}